\newif\ifemulate
\emulatetrue

\ifemulate
  \documentclass{emulateapj}
  \usepackage{apjfonts}
\else
\fi

\newcommand{\etal}{et~al.}

\def\figone{
\begin{figure*}
\begin{center}
  \includegraphics[width=0.90\textwidth]{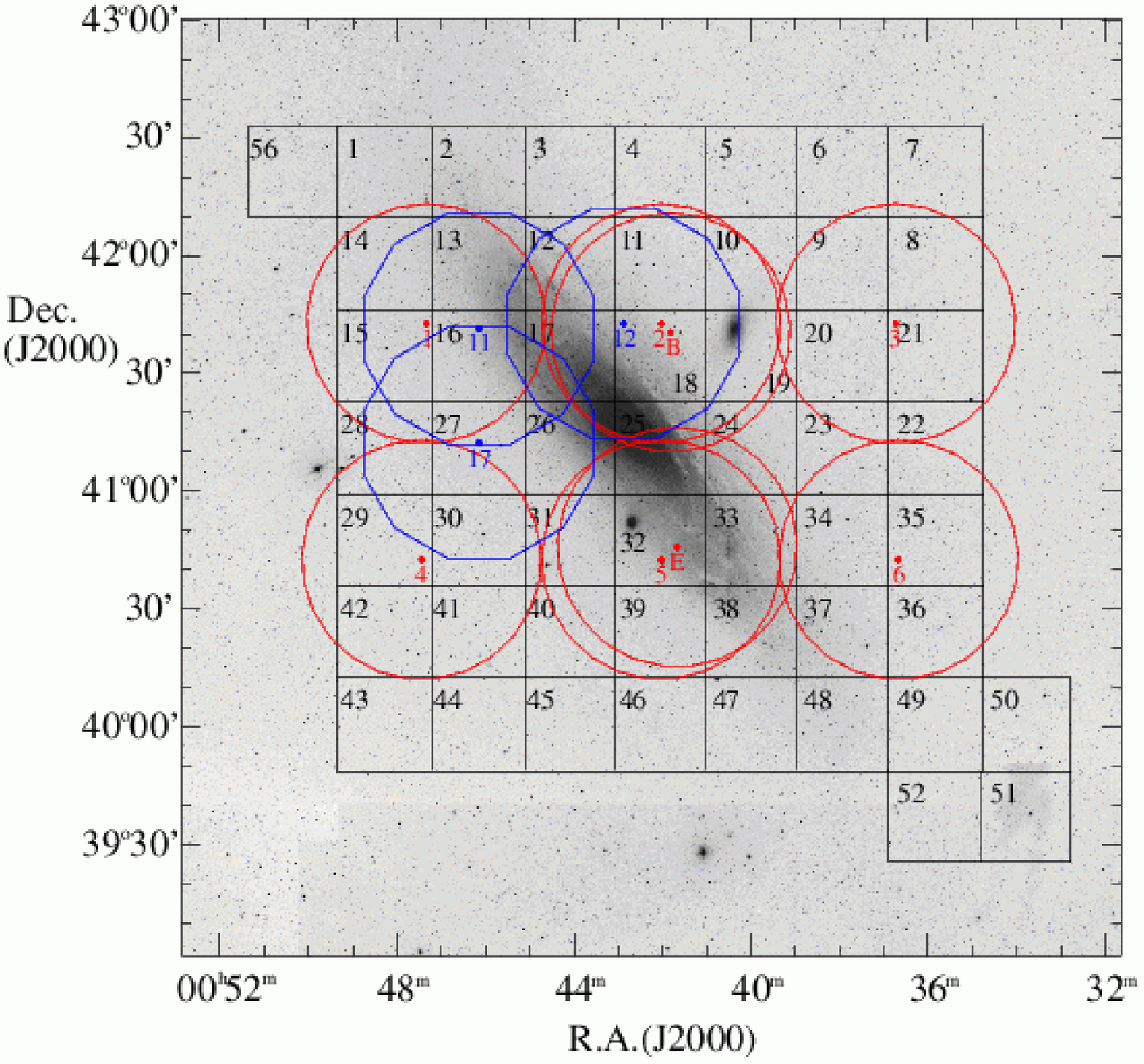} 
\end{center}
\caption{
Locations of our survey regions overlaied on a $4\arcdeg \times 4\arcdeg$ 
  optical image of M31 from the Digitized Sky Survey 
(image center : R.A.(J2000.0)$=0^h~ 42^m~ 15^s$, Dec. (J2000.0)$= +41\arcdeg~ 00'~ 37''$).
North is at the top and east is to the left.
The fields for the KPNO 0.9 m photometric imaging
  are drawn with solid boxes 
  with the field names labeled in the upper left corner of each box.
The field of view of each field is $23.'2 \times 23.'2$
  and the observations were made so that there was some ($1' - 2'$) overlap
  between adjacent fields.
Satellite galaxies NGC 205 and M32 are located in fields 19 and 32, respectively.
The Hydra fields 
  are denoted by smooth circles (for 2000) and polygon-like circles (for 2001),
  and their names are labelled near the center of each field.
}
\end{figure*}
}

\def\figtwo{
\begin{figure}
\includegraphics[width=0.47\textwidth]{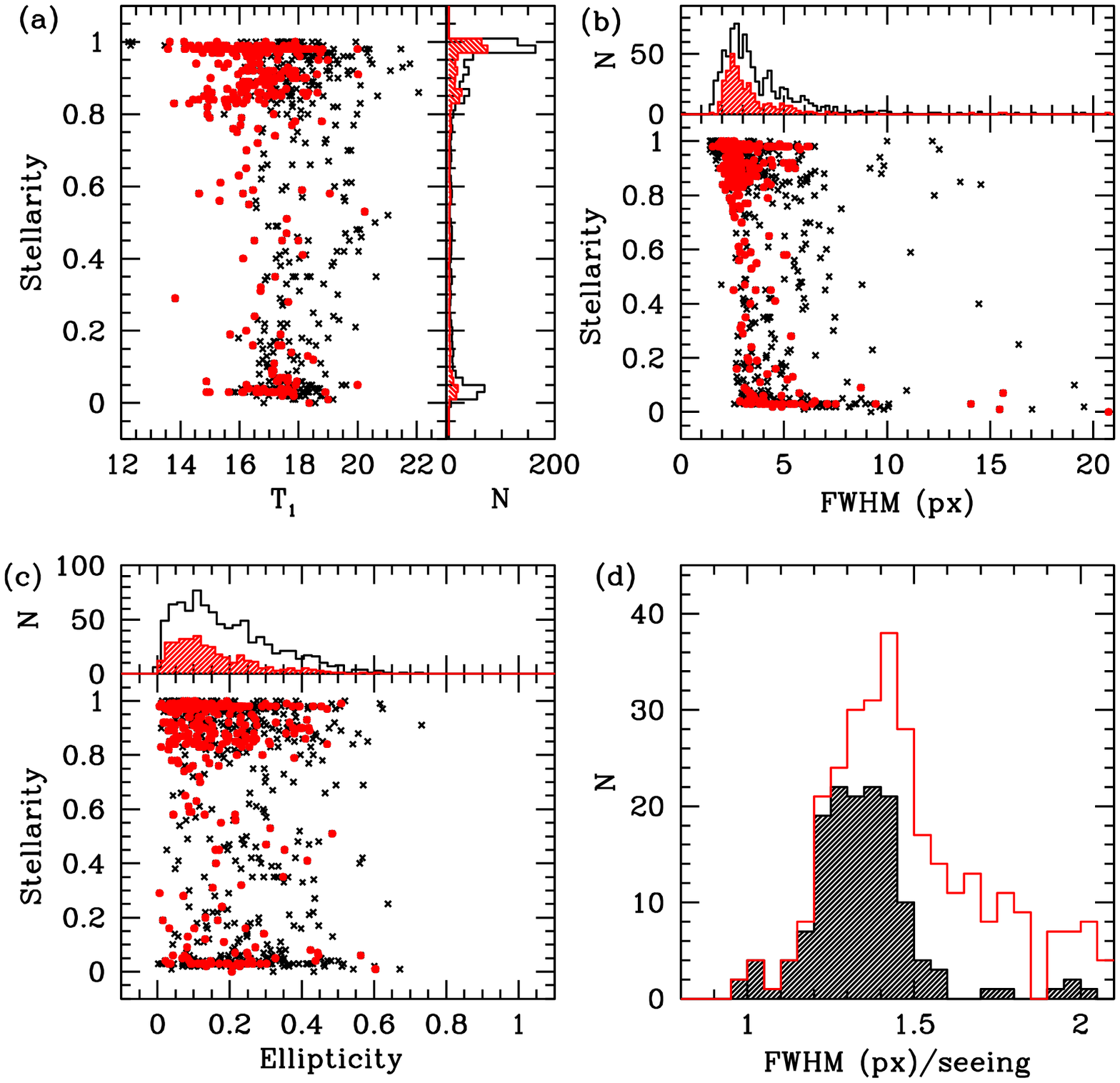} 
\caption{
SExtractor parameter distributions of the 347 confirmed and 514 candidate GCs
  in the catalogs of \citet[RBC2]{gal06},
  \citet{hux05}, and \citet{mac07} matched with our photometry.
(a) Stellarity vs. $T_1$ magnitude; (b) stellarity vs. FWHM;
  (c) stellarity vs. ellipticity; 
  (d) histogram of FWHM/seeing distribution.
In panels (a), (b), and (c),
  the crosses and the open solid histograms show
  the distributions of all 861 objects with good photometry,
  and the filled circles and hatched histograms show
  those of the 347 confirmed GCs.
In panel (d), 
  the open solid histogram shows the distribution of 
    the 347 confirmed GCs and 
  the hatched histogram the 147 confirmed compact GCs with stellarity $>0.95$.
}
\end{figure}
}

\def\figthree{
\begin{figure}
\includegraphics[width=0.45\textwidth]{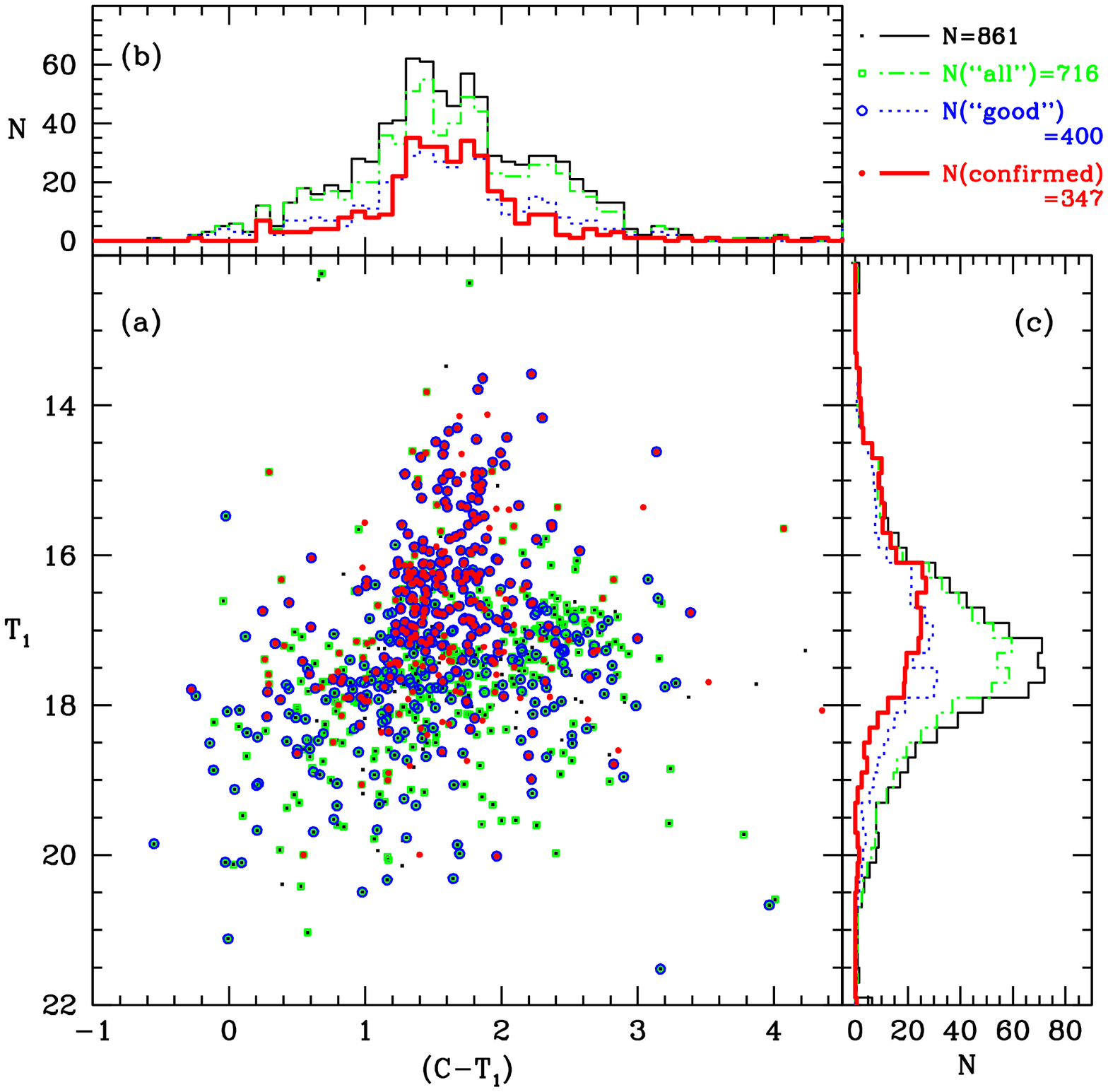} 
\caption{
(a) Color-magnitude diagram; (b) $(C-T_1)$ color distribution; 
  (c) $T_1$ LF of the 347 confirmed and 514 candidate GCs
  in the catalogs of \citet[RBC2]{gal06},
  \citet{hux05}, and \citet{mac07} matched with our photometry.
Small dots and thin solid histograms show the distributions of
  all 861 objects with good photometry,
 open squares and dot-dashed histograms show the distributions 
  of ``all GC candidates'' from the GC search criterion 1,
 open circles and dotted histograms show the distributions 
  of ``good GC candidates'' from criterion 2, and
 filled circles and thick solid histograms show the distributions 
  of the confirmed GCs in the three papers above.
\label{fig3}}
\end{figure}
}

\def\figfour{
\begin{figure}
\includegraphics[width=0.47\textwidth]{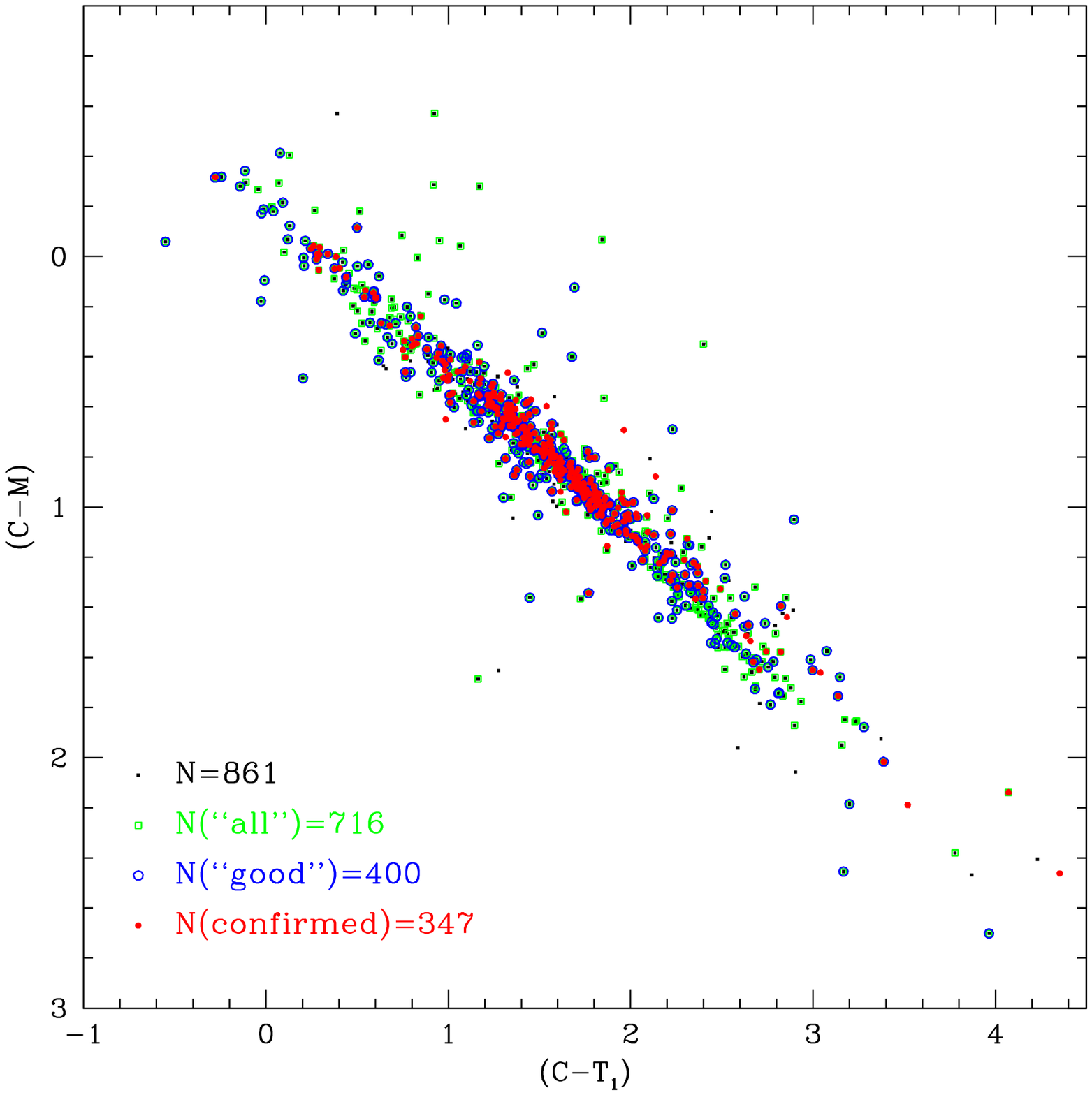}
\caption{
Color-color diagram
  of the 347 confirmed and 514 candidate GCs
  in the catalogs of \citet[RBC2]{gal06},
  \citet{hux05}, and \citet{mac07} matched with our photometry.
 Small dots show the distributions of all 861 objects with good photometry,
 open squares show the distributions
  of ``all GC candidates'' from the GC search criterion 1,
 open circles show the distributions
  of ``good GC candidates'' from criterion 2, and
 filled circles show the distributions
  of the confirmed GCs in the three papers above.
\label{fig4}}
\end{figure}
}

\def\figfive{
\begin{figure*}
\includegraphics[width=0.90\textwidth]{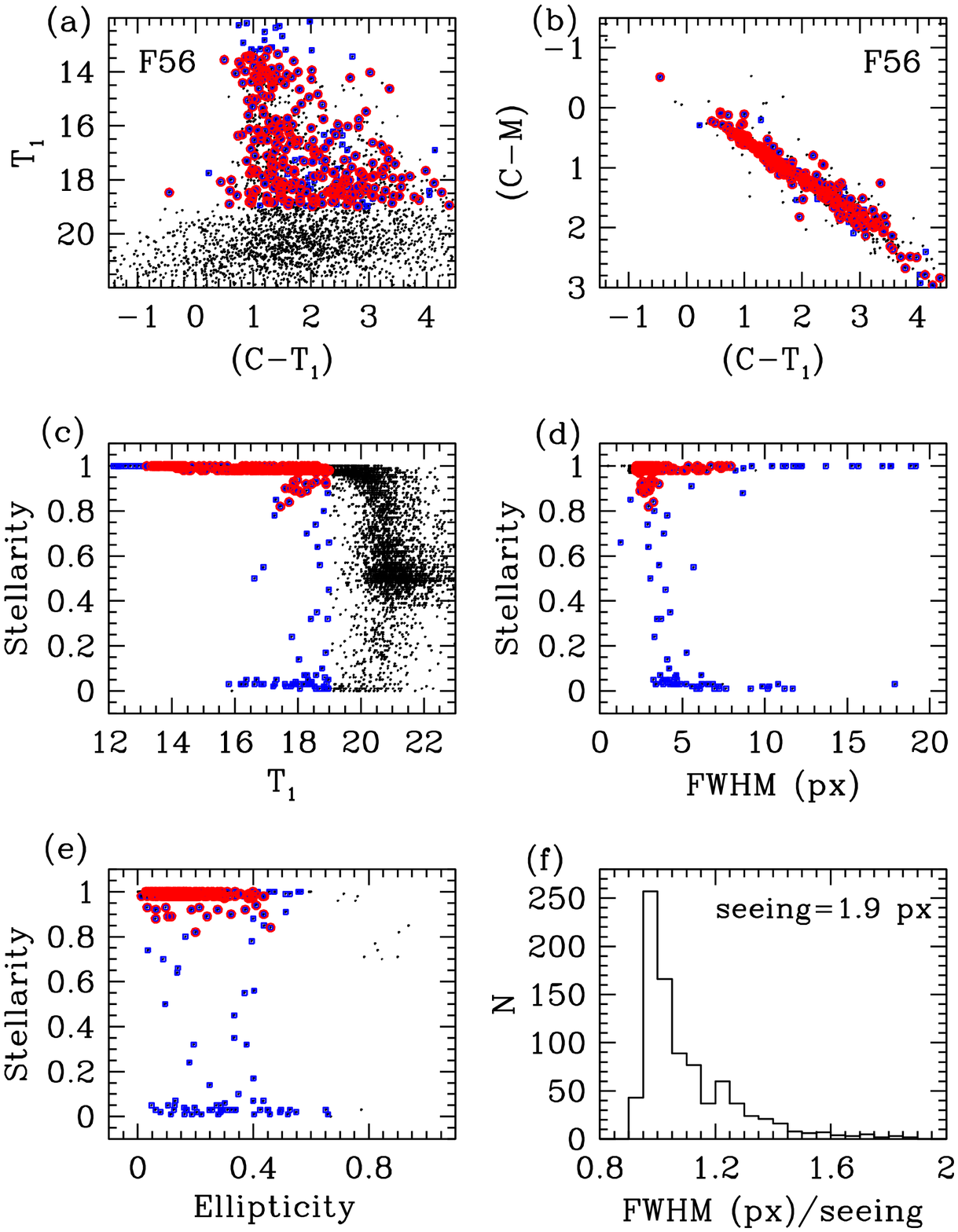} 
\caption{
Example of our application of the GC search criteria 
  to one of the KPNO fields (F56)
  with seeing of $\approx 1.9$ pixels.
Panels (a) and (b) show the CM diagram and CC diagram 
  of the selected GC candidates, respectively, 
  and panels (c) -- (f) show the parameter space 
  for the GC candidate selection (see Section 3 for details).
Small dots represent all measured objects with good photometry (N=4894),
  squares are ``all GC candidates'' selected 
   according to criterion 1 (N=362), and
 open circles superimposed on the squares represent ``good GC candidates'' selected
   according to criterion 2 (N=277).
}
\end{figure*}
}

\def\figsix{
\begin{figure*}
\begin{center}
  \includegraphics[width=0.80\textwidth]{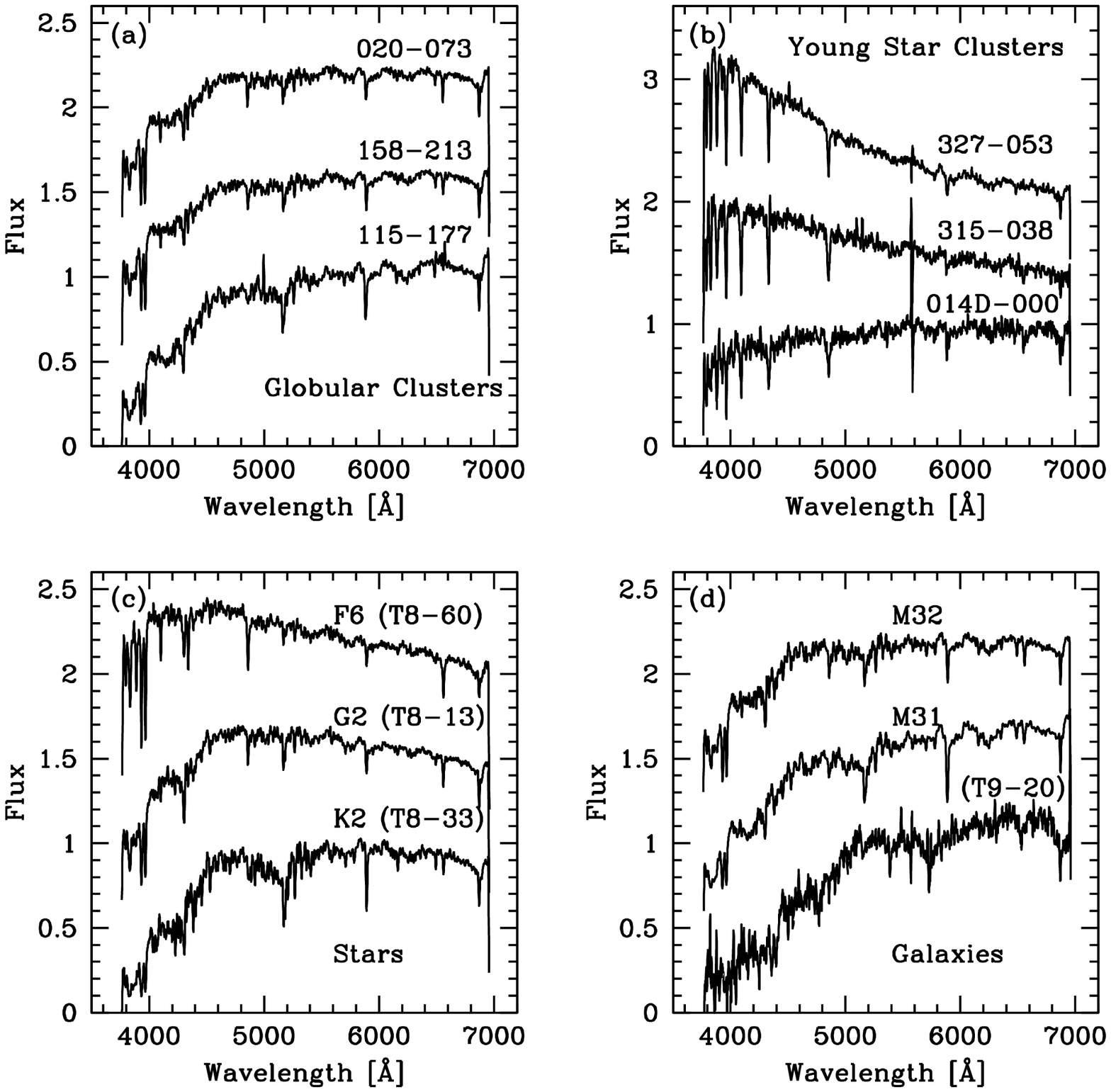} 
\end{center}
\caption{
Sample spectra of (a) M31 globular clusters,
  (b) M31 young star clusters, (c) foreground stars,
  and (d) three galaxies (M31, M32 and a background galaxy).
We have followed \citet{huc91} and \citet{bar00} for the naming convention
  of M31 GCs.
The numbers after T8$-$ and T9$-$ in the parentheses of panels (c) and (d) are
  the identification numbers in Tables $16-17$ and 18, respectively.
\label{fig6}}
\end{figure*}
}

\def\figseven{
\begin{figure*}
\includegraphics[width=0.90\textwidth]{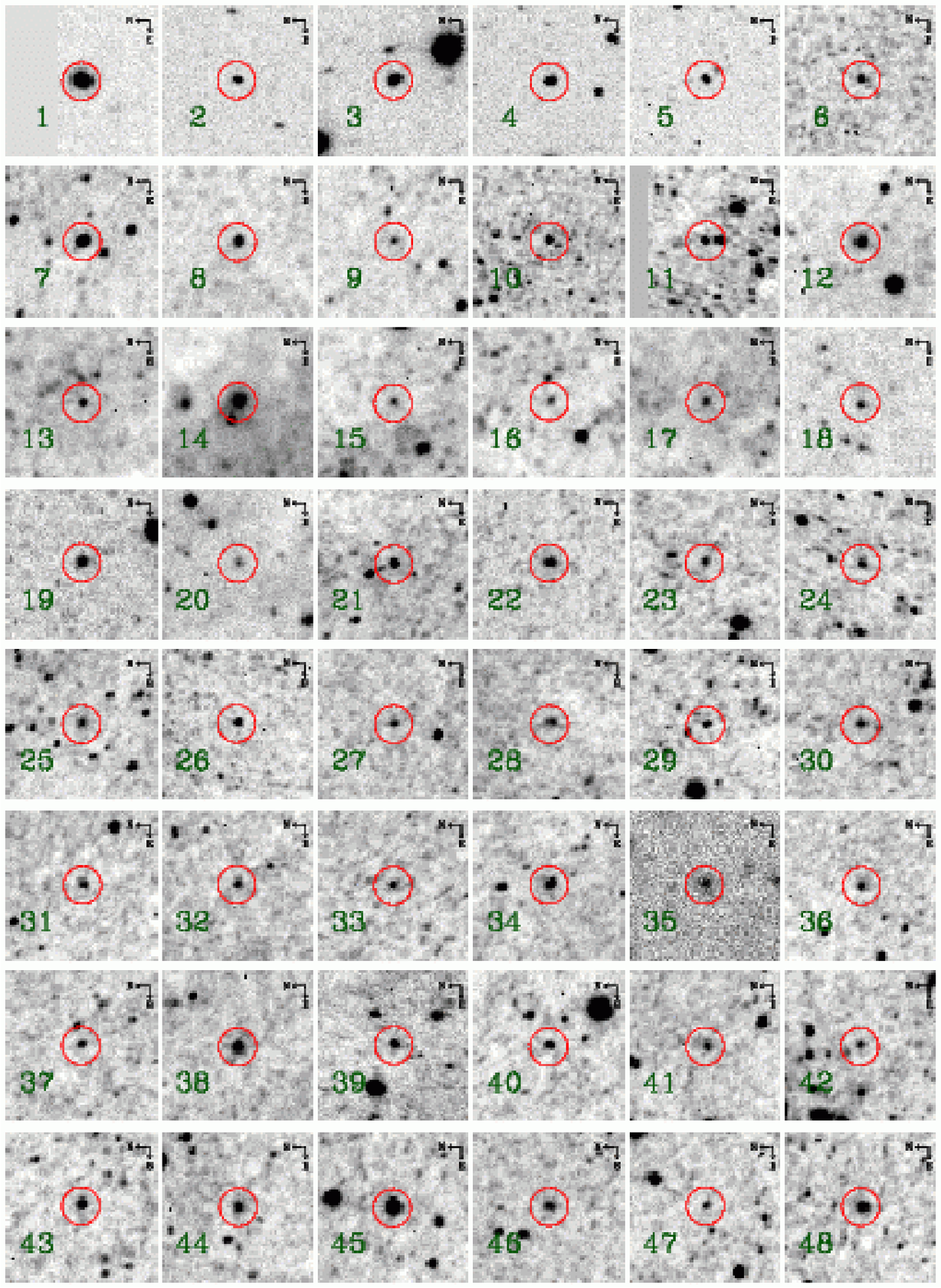} 
\caption{
Gray-scale mosaic maps of the $R$-band CCD images of genuine GCs (class 1)
  for identifications from 1 to 48 in Tables $5-6$.
The size of each field is $40\arcsec \times 40\arcsec $,
  with north to the left and east at the bottom.
The GC is centered in each image, in the center of the circle of 
  $5\arcsec$ radius.
}
\end{figure*} }

\def\figeight{
\begin{figure*}
\includegraphics[width=0.90\textwidth]{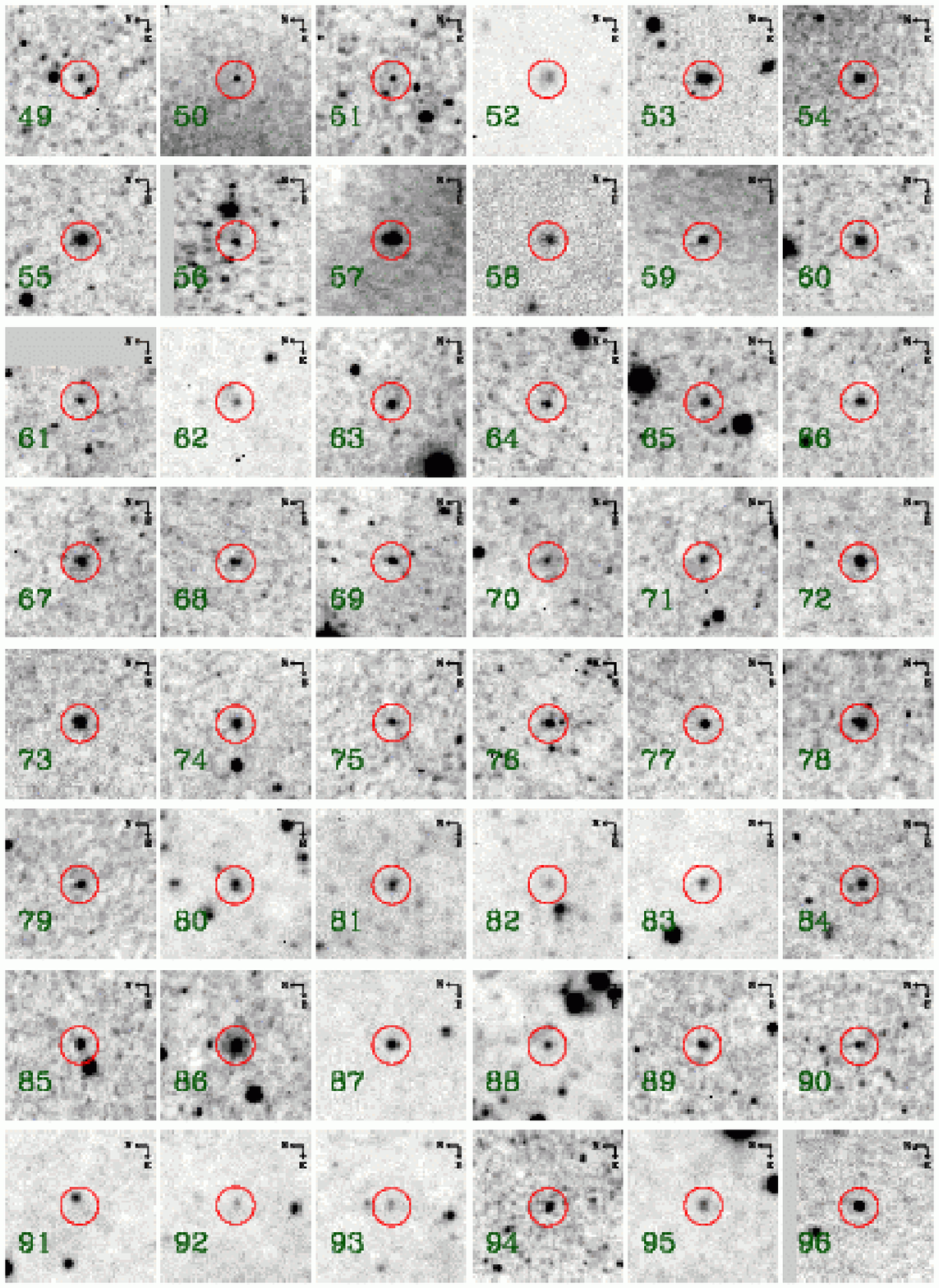} 
\caption{
Same as Figure 7, but for identifications from 49 to 96.
}
\end{figure*} }

\def\fignine{
\begin{figure*}
\includegraphics[width=0.90\textwidth]{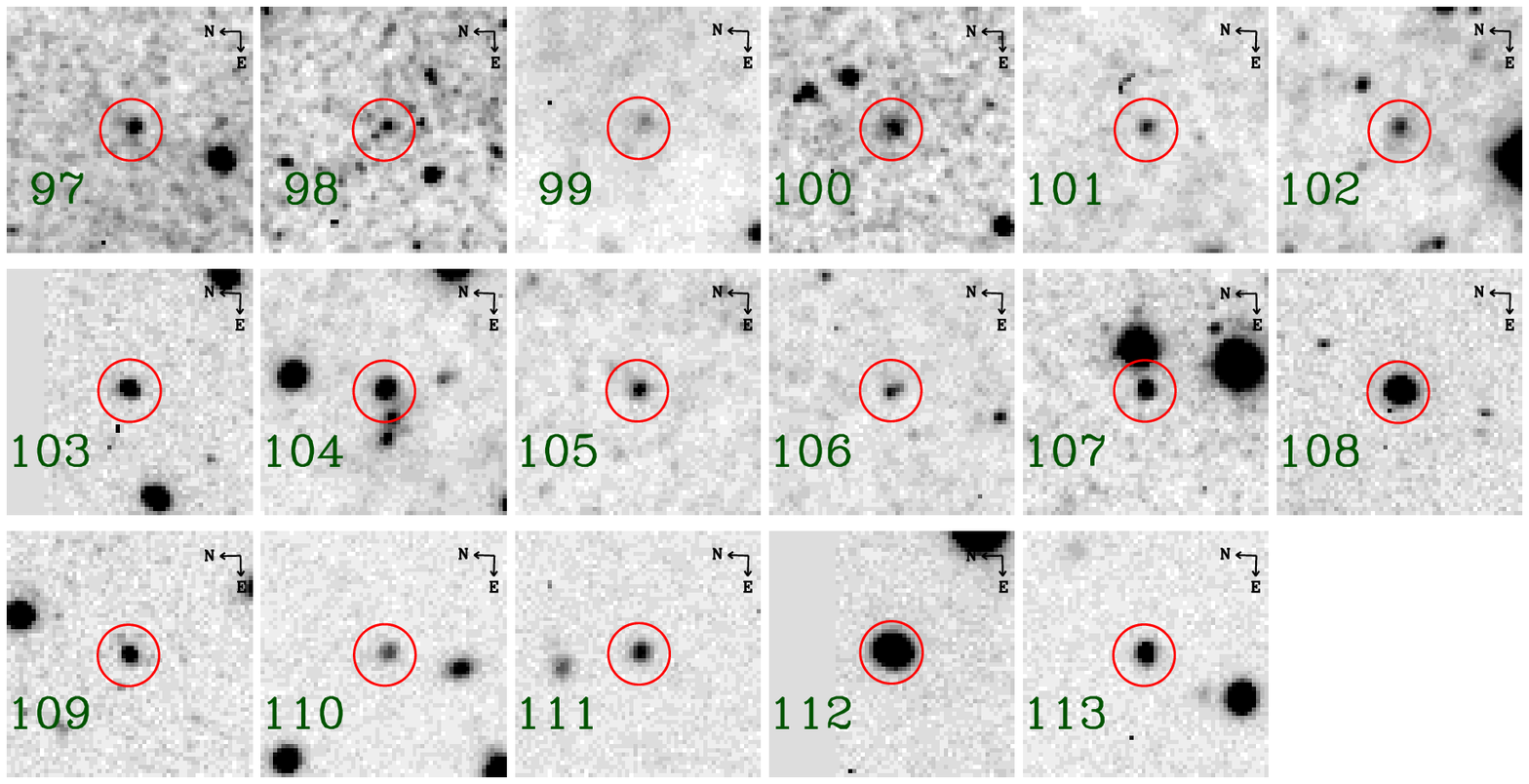} 
\caption{
Same as Figure 7, but for identifications from 97 to 113.
}
\end{figure*} }

\def\figten{
\begin{figure*}
\includegraphics[width=0.90\textwidth]{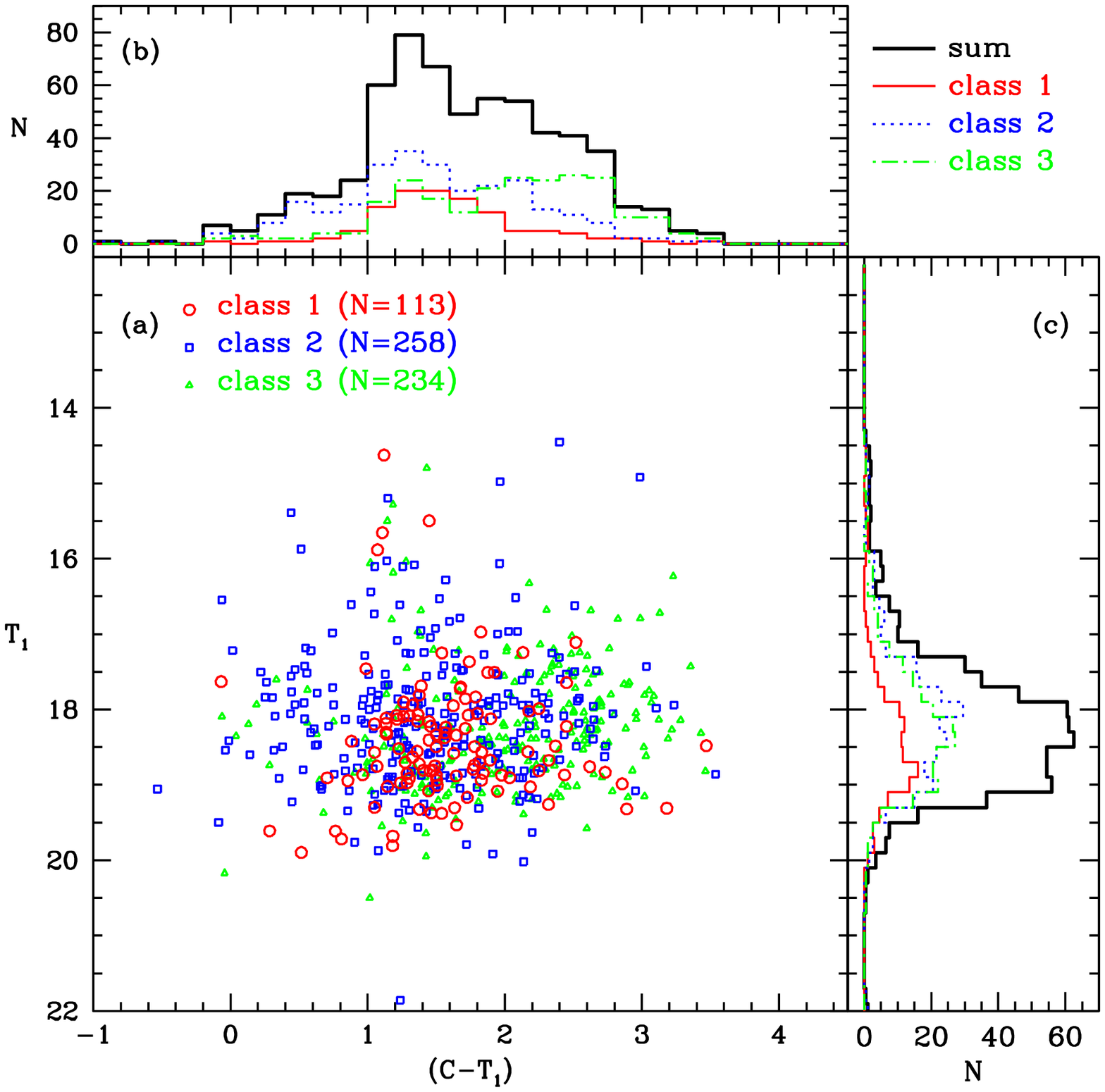} 
\caption{
Photometric diagrams of the three classes of GCs/GC candidates newly found 
  in this study. 
(a) $T_1-(C-T_1)$ diagram; 
  (b) $(C-T_1)$ color distribution;
  (c) $T_1$ LF.
Open circles and thin solid histograms are for genuine GCs (class 1),
  open squares and dotted histograms are for probable GCs (class 2),
  and open triangles and dot-dashed histograms are for possible GCs (class 3).
Thick solid histograms in panels (b) and (c) are the sum of 
  the numbers of all three classes.
}
\end{figure*}
}

\def\figeleven{
\begin{figure}
\begin{center}
   \includegraphics[width=0.47\textwidth]{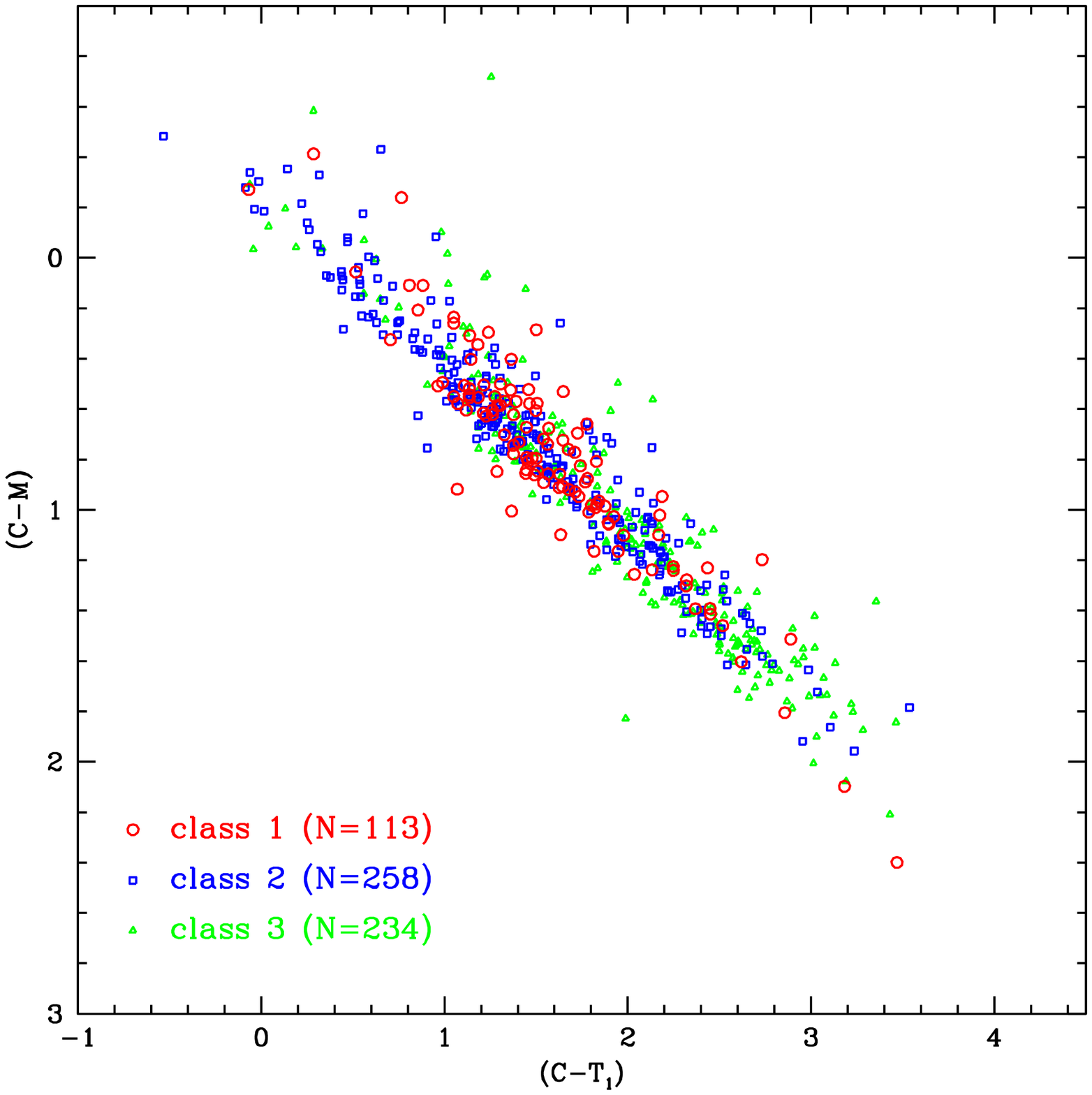}
\end{center}
\caption{
$(C-M)-(C-T_1)$ diagram of the three classes of GCs/GC candidates newly found
  in this study.
Open circles are for genuine GCs (class 1),
  open squares are for probable GCs (class 2),
  and open triangles are for possible GCs (class 3).
}
\end{figure} }

\def\figtwelve{
\begin{figure}
\begin{center}
  \includegraphics[height=0.47\textwidth]{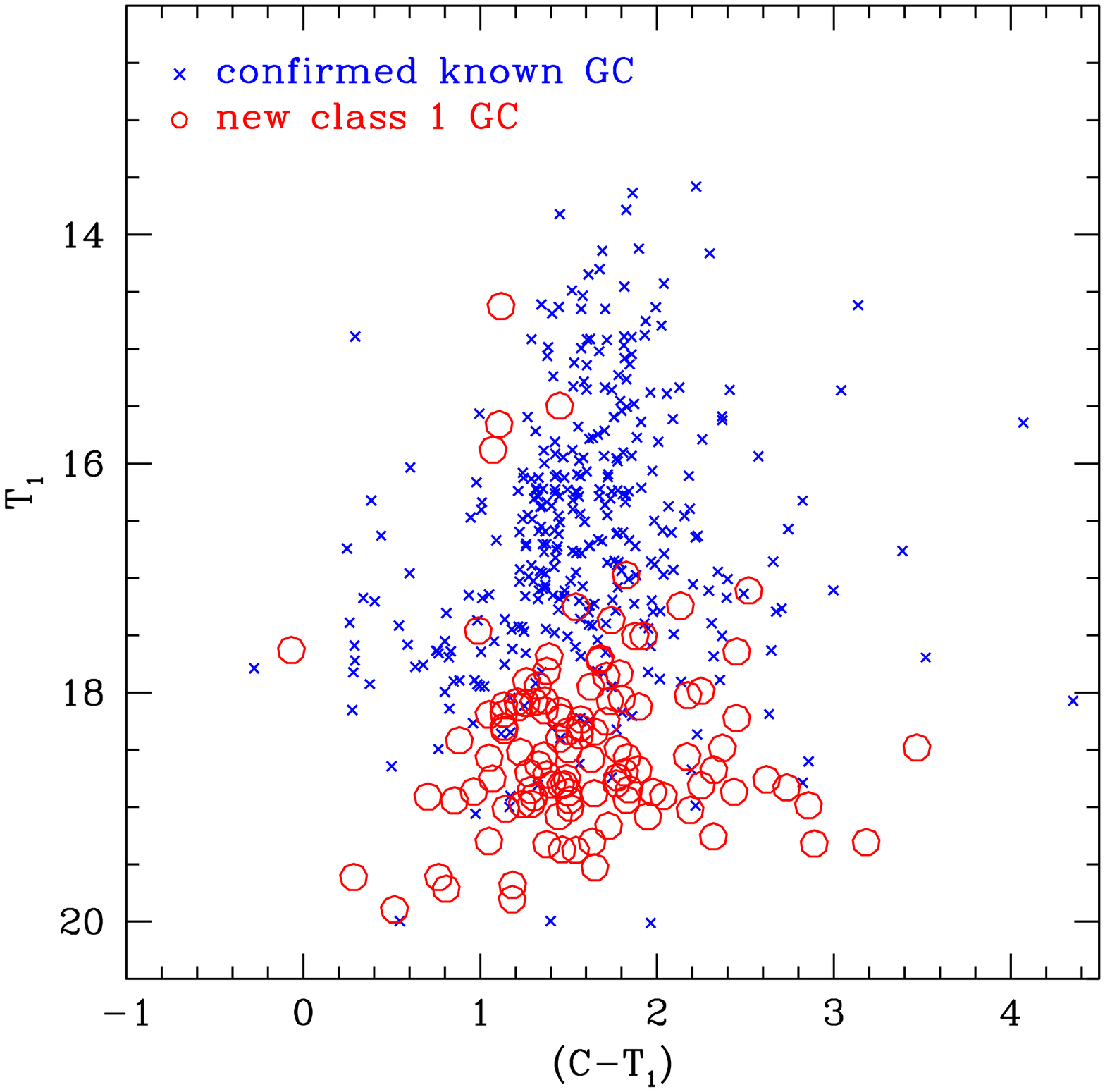}
\end{center}
\caption{
$T_1-(C-T_1)$ CM diagram of confirmed, previously known GCs 
  in Figure 3 (a) (crosses) and
  newly found, class 1 GCs in Figure 10 (a) (open circles).
}
\end{figure}
}

\def\figthirteen{
\begin{figure}
\includegraphics[width=0.47\textwidth]{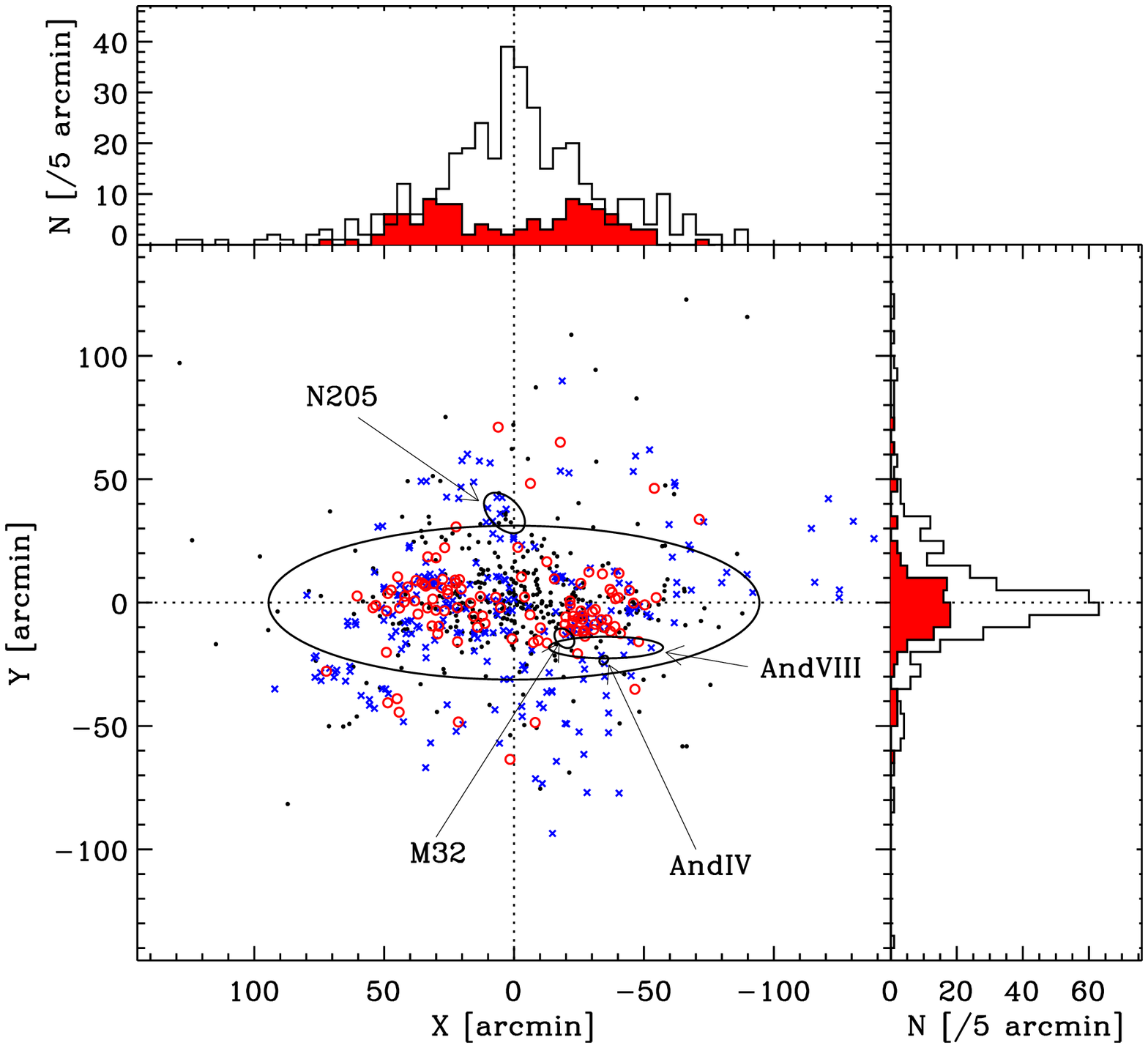}
\caption{
Spatial distribution and histograms of the newly found GCs and GC candidates.
$X$ is the distance measured parallel to the major axis of M31
  increasing to the north-east direction, and $Y$ is the
  distance parallel to the minor axis.
The large ellipse is for M31 with
  position angle 37.7\arcdeg \citep{racine91}.
The positional data of the galaxies are from \citet{kara04}, 
  while those for And VIII are from \citet{morrison03}.
Open circles are for class 1 GCs in this study,
  crosses are for class 2 GCs in this study, and
  dots are for GCs in 
  the catalogs of \citet[their class 1]{gal06},
  \citet{hux05}, and \citet{mac07}.
The filled histograms are for class 1 GCs in this study, and 
  the open solid histograms are for GCs in 
  the three papers above.
\label{fig13}}
\end{figure}
}

\shorttitle{Globular Clusters in M31. I.}
\shortauthors{Kim et al.}

\begin{document}

\title{Wide-Field Survey of Globular Clusters in M31. I. A Catalog of New Clusters}

\author{Sang Chul Kim\altaffilmark{1,2,3}, 
Myung Gyoon Lee\altaffilmark{2,3},
Doug Geisler\altaffilmark{3,4}, 
Ata Sarajedini\altaffilmark{3,5}, 
Hong Soo Park\altaffilmark{2},
Ho Seong Hwang\altaffilmark{2},
William E. Harris\altaffilmark{6},
Juan C. Seguel\altaffilmark{4,7}, \&
Ted von Hippel\altaffilmark{3,8,9}}

\altaffiltext{1}{Korea Astronomy and Space Science Institute, Daejeon 305-348, 
   Korea; sckim@kasi.re.kr}
\altaffiltext{2}{Astronomy Program, Department of Physics and Astronomy, Seoul National University,
   Seoul 151-742, Korea; mglee@astrog.snu.ac.kr, hspark@astro.snu.ac.kr, hshwang@astro.snu.ac.kr}
\altaffiltext{3}{Visiting Astronomer, Kitt Peak National Observatory, 
National Optical Astronomy Observatory, which is operated by the Association 
of Universities for Research in Astronomy (AURA), Inc., under cooperative 
agreement with the National Science Foundation.}
\altaffiltext{4}{Grupo de Astronomia, Departamento de Fisica, Universidad de
   Concepci\'{o}n, Casilla 160-C, Concepci\'{o}n, Chile; dgeisler@astro-udec.cl,
   jseguel@andromeda.cfm.udec.cl}
\altaffiltext{5}{Department of Astronomy, University of Florida, 
   Gainesville, FL 32611, USA; ata@astro.ufl.edu}
\altaffiltext{6}{Department of Physics and Astronomy, McMaster University, 
   Hamilton, ON L8S 4M1, Canada; harris@physics.mcmaster.ca}
\altaffiltext{7}{Cerro Tololo Inter-American Observatory, Casilla 603, La Serena, Chile}
\altaffiltext{8}{Department of Astronomy, University of Texas at Austin, 
   1 University Station C1400, Austin, TX 78712, USA; ted@astro.as.utexas.edu}
\altaffiltext{9}{Visiting Scientist, Southwest Research Institute,
   1050 Walnut Street, Suite 400, Boulder, CO 80302, USA}

\begin{abstract}

We present the result of a wide-field survey of globular clusters (GCs) 
  in M31 covering a $3\arcdeg \times 3\arcdeg$ field centered on M31.
We have searched for GCs on CCD images taken with Washington $CMT_1$ filters 
  at the KPNO 0.9 m telescope using the following steps:
(1) inspection of morphological parameters given by the SExtractor package 
  such as stellarity, full width at half-maximum, and ellipticity;
(2) consulting the spectral types and radial velocities obtained from spectra
  taken with the Hydra spectrograph at the WIYN 3.5 m telescope; and
(3) visual inspection of the images of each object.
We have found 1164 GCs and GC candidates,
  of which 605 are newly found GCs and GC candidates
  and 559 are previously known GCs.
Among the new objects there are 113 genuine GCs, 258 probable GCs, and 234 possible GCs,
  according to our classification criteria.
Among the known objects there are 383 genuine GCs, 109 probable GCs, 
  and 67 possible GCs.
In total there are 496 genuine GCs, 367 probable GCs and 301 possible GCs.
Most of these newly found GCs have $T_1$ magnitudes of $17.5 - 19.5$ mag,
[$17.9 < V < 19.9$ mag assuming $(C-T_1) \approx 1.5$],
and $(C-T_1)$ colors in the range $1 - 2$.
\end{abstract}

\keywords{galaxies: star clusters --- galaxies: spiral ---
galaxies: individual (M31, NGC 224) --- Local Group}

\section{Introduction}

Globular clusters (GCs)
 are an ideal tool for studying the formation and evolution of nearby galaxies 
for several reasons.
First, GCs are one of the brightest objects in galaxies,
  so it is relatively easy to observe them 
  even in the outer parts of the galaxies where individual stars are too faint to be observed.
Second, GCs are believed to be among the oldest objects in galaxies 
  (see, e.g., \citet{sal02, dea05}) 
   giving a lower limit to the ages of their parent galaxies.
Third, the stars in GCs are believed to be born essentially at the same time
  and with the same chemical composition,
  which makes GCs an ideal laboratory for the study of stellar evolution.
Fourth, GCs are distributed much more widely than stars, so they can be
  used for the study of the halo of their parent galaxy.
Finally, since the present GCs have survived since the formation of their parent galaxies,
  they give information on the formation and evolution of both the clusters and the galaxies.

The GCs in M31 are especially  important, 
  since M31 is the nearest spiral galaxy and has an abundant population of GCs.
There have been numerous studies of the GCs in M31 starting as early as 1932. 
Table 1 shows a list of the previous studies on M31 GCs, focusing primarily on
  the number of GCs and candidate GCs found.
Examples of the most extensive GC surveys are those of \citet[the M31 Consortium]{sar77},
  \citet[the DAO group]{cra85},
  and \citet[the Bologna group]{bat87}.
However, these surveys are mostly 
based on visual searches of the photographic plates.

Since the use of CCD detectors in astronomy, there have been efforts to use them
  for deep photometry to search for new GCs in M31.
However, the small field of view (FOV) of the first generation CCDs enabled
  previous investigators only to perform GC surveys for a limited region of M31,
  and there has not yet been any wide-field survey of GCs using CCD cameras.
It is clear from Table 1 that our new GC survey presented in this study
  is the first systematic one for the largest area of 
  $\sim$ 3\arcdeg $\times$ 3\arcdeg~ centered on M31.

Recently several new extended GCs were found in the halo located 
  at $15 \lesssim R_p \lesssim 116$ kpc (where $R_p$ is the projected radius)
  from the center of M31 \citep{hux05, mar06, mac07}, and 
  \citet{mac06, mac07} presented deep photometry of stars
  in these clusters based on {\it Hubble Space Telescope (HST)} ACS images.
\citet{kod04} found 49 compact star clusters 
  with $M_V<-5$ mag and $0<(B-V)<1.0$ in the south-west field 
  ($17'.5 \times 28'.5$) of the M31 disk from CCD images 
  taken at the Subaru 8 m telescope, some of which may be GCs.

This paper is the first in a series on our wide-field survey of M31 GCs.
In this paper we present a catalog of new GCs in M31, and 
  the analyses of the photometric and spectroscopic data of 
  the new and known GCs in M31 will be presented in separate papers.
Brief progress reports of this study were given in \citet{lee02},
  \citet{kim02}, and \citet{seg02},
  which are superseded by this series of papers.

This paper is organized as follows:
\S 2 describes the photometric and spectroscopic observations and
data reductions, and \S 3 the GC search method.
Section 4 presents the catalog of new GCs found in this study and
  some properties of newly found GCs,
  and finally, a summary is given in \S 5.

\section{Observations and Data Reduction}

We carried out two kinds of observation for the survey of M31 GCs.
First, photometric observations were made using the CCD camera at the KPNO 0.9 m telescope.
Second, spectroscopic observations were performed using the Hydra multifiber spectrograph
  at the WIYN 3.5 m telescope. 
We describe the details of these observations below.

\subsection{Photometry}
   \subsubsection{Observation}

\ifemulate\figone\fi

We obtained Washington $C$ and $M$ and broadband Kron-Cousins $R$ images 
  using the T2KA CCD camera at the KPNO 0.9 m telescope
  on the nights of UT 1996 October 14 -- 25 and UT 1998 October 19.
The pixel scale of the CCD chip is $0.68''$ pixel$^{-1}$, and
  the CCD has 2048 $\times$ 2048 pixels, corresponding to 
  $23.2' \times 23.2'$ on the sky.
We used the Kron-Cousins $R$ filter as an alternative to the $T_1$ filter,
  since the $R$ filter accurately reproduces the $T_1$ photometry
  with 3 times greater efficiency \citep{gei96}.
The resulting calibrated magnitudes and colors will therefore be
  in the Washington $CMT_1$ system.
\citet{gei96}
  gave a transformation relation between $R$ and $T_1$,
  $R = 0.003 + T_1  - 0.017 \* (C-T_1)$,
  with an rms of only 0.02 mag derived from the data of 53 standard stars.

We observed 53 fields
  covering the central region of M31.
Figure 1 shows the location of the observed fields,
  the names of which are labeled in the upper left corner of each solid box.
For most of the fields, one exposure per filter was made.
Typical exposure times for the 1996 run were 1500 s for $C$ and 600 s for $M$ and $R$,
  while those for the 1998 run were 1200 s for $C$ and 500 s for $M$ and $R$.
The seeing was mostly 1.1\arcsec -- 2.0\arcsec (1.6 -- 2.9 pixels in our CCD frames)
  during the observations,
  although a few fields have a seeing of 2.0\arcsec -- 3.0\arcsec.
Table 2 lists the journal of observation, where 
  column (1) is the night number, column (2) the observation date in UT,
  column (3) the field numbers, and column (4) the weather condition.
Of the total of 11 nights of observation, 
  four nights (N5, N8, N9 and Oct98) were photometric,
  three (N3, N4, and N11) were semi-photometric,
  and the remaining four nights were non-photometric or even cloudy.
The standard star observations were made in the photometric and semi-photometric nights
  indicated in Table 2.

\subsubsection{Data Reduction}

We processed all the CCD images to apply
  overscan correction, bias subtraction and flat fielding using the 
  IRAF\footnote{IRAF(Image Reduction and Analysis Facility) is distributed 
  by the National Optical Astronomy Observatory, which is operated 
  by AURA, Inc., 
  under cooperative agreement with the National Science Foundation.}/{\small CCDRED} package.
We derived the calibration transformation using 
  the Washington standard stars \citep{gei96}
  observed during the observing runs.
We obtained the aperture magnitudes of the standard stars using a $7.5''$ radius aperture 
  (the same as in \citet{gei96}) from the images of the standard stars.
Then we used the IRAF {\small PHOTCAL} package to derive the calibration equations.

For three out of the four nights of photometric conditions,
  all three standard calibration coefficients (zeropoint, color term, and airmass term)
  were derived.
For the three nights of semi-photometric conditions,
  we adopted the mean values of the color and airmass term coefficients of
  the three photometric nights
  to derive the zeropoints.
Although the night of UT 1996 October 22 was believed to be photometric,
  there were not enough standard stars to derive
  all three calibration coefficients independently,
so we adopted the color and airmass term coefficients of the previous photometric night.

For the fields observed on the nights without standard stars,
  we derived secondary standard transformations using the neighboring fields.
Since we initially arranged our target fields to overlap 
  adjacent fields by $1'$ -- $2'$,
  we could easily identify the stars in common between two neighboring fields.
On the image of each filter, we first identified the positions of the
  common stars in the two adjacent fields,
  calculated the mean magnitude offsets
  between the standardized magnitudes and instrumental magnitudes, and then
  applied this magnitude offset (together with the color and atmospheric
  coefficients of the standardized field) to transform the instrumental magnitudes
  of the nonphotometric frames.
There are 14 fields for which this secondary transformation method is applied.
The typical errors of the standard star calibration are 
  0.020, 0.022, and 0.019 mag for $T_1$, $(C-T_1)$, and $(M-T_1)$, respectively.

We have derived the photometry of the objects in the target images using
  the SExtractor package \citep{ber96}.
SExtractor performs detection of objects in the images and gives
  position, aperture magnitude, stellarity, the full width at half-maximum (FWHM), 
  ellipticity, position angle, quality of the photometry, and some other parameters. 
We used the SExtractor parameters DETECT\_MINAREA = 5 pixel and
  DETECT\_THRESH = $1.5\sigma$ above the local background.
The results for new GC searches do not depend strongly on the choice of these values.
The instrumental magnitudes of the objects obtained using the SExtractor package
  were transformed into the standard system using the 
  calibration equations.

We have obtained plate solutions for each of the CCD images 
  for astrometry of objects
  using the Guide Star Catalogue (GSC) provided by the Space Telescope Science
  Institute (STScI) and the IRAF tasks {\sc ccxymatch} and {\sc ccmap}.
These plate solutions transform the $X$ and $Y$ coordinates of our $R$ images
  to/from the celestial equatorial coordinates of epoch J2000.0
  by using the IRAF {\sc cctran} task.
The mean rms errors in right ascension and declination are $0.064'' \pm 0.023''$ and
  $0.064'' \pm 0.024''$, respectively.

   \subsection{Spectroscopy}

   \subsubsection{Observation}

For most of the GC candidates selected from the photometric list of the objects to
  be described in the next section,
  we carried out spectroscopic observation 
  using the Hydra multifiber bench spectrograph and T2KC CCD
  at the WIYN 3.5 m telescope
  on the nights of UT 2000 September 7--9 and UT 2001 November 2--4.
Table 3 lists the journal of spectroscopic observations.
Table 3 shows the observation date in UT, the number of target objects, 
  and the exposure time for each Hydra configuration marked in Figure 1.

For both the 2000 and 2001 observations, almost the same instrumental setup was used.
The 400@4.2 grating and Simmons camera were used.
This combination with the blue fiber cable 
  covers a wavelength range of $\sim 3400 - 6600$ \AA~ in the first order and
  gives a 7.07 \AA~ spectral resolution and 1.56 \AA ~pixel$^{-1}$ dispersion.

During the observing run of 2000, all three nights were clear and a total of eight Hydra
  configurations were used for spectroscopy.
However, in the run of 2001 only the first
  night was clear, while the two subsequent nights were cloudy or rainy.
Only three Hydra configurations were obtained during this run.
The total number of targets observed during the observing runs
  was 748, including 106 previously known GCs observed in order to quantify
  our errors and compare our values with previous studies.

\subsubsection{Data Reduction}

First we performed overscan correction, image trimming, bias subtraction, 
  and flat combining on the spectroscopic data
  using the IRAF {\sc ccdred} package. 
We removed cosmic rays in the object images and combined the resulting
  object images.
For the reduction of the Hydra spectroscopic data,
  we used the Hydra data reduction task, IRAF {\sc dohydra},
  which was specifically designed for multifiber spectral reduction \citep{val95}.

Before doing the main data reduction part,
  DOHYDRA first performs aperture finding using the {\sc apfind} task,
  and performs fitting and subtracting of the scattered light using the
  {\sc apscatter} task.
Then DOHYDRA performs
  aperture extraction, 
  flat-fielding, fiber throughput correction, wavelength calibration, 
  and sky subtraction.
Dome-flat images were used as a template to extract the one-dimensional object
  and calibration spectra from the two-dimensional images.
Cu/Ar calibration lamp spectra were used for wavelength calibration.
The rms error of the wavelength calibration is estimated to be typically 
0.2--0.3 \AA.
Finally, we calibrated the flux of the spectra of the targets using the spectra
  of the flux standard star BD +40 4032 (R.A.(B1950)$=20^h~ 06^m~ 40^s.0$, 
  Dec.(B1950)$= +41\arcdeg~ 06'~ 15''$, B2 III, m$_{5556}=10.45$ mag; 
  \citet{strom77}) using the {\sc calibrate} task.

We determined the radial velocity of the targets 
  by cross-correlating their spectra
  against high signal-to-noise ratio (S/N) template spectra 
  using the IRAF {\sc fxcor} task \citep{ton79, huc91}.
We used two bright GCs in M31 as a reference.
GCs 020-073 and 158-213 are relatively bright clusters 
  with $V=14.91$ mag and $(B-V)=0.83$ and $V=14.70$ mag and $(B-V)=0.86$, respectively,
  and have well-determined radial velocities of 
  $V_r = -349 \pm 2$ and $-183 \pm 4$ km s$^{-1}$, respectively
  \citep{bar00}.
We used the wavelength range of 3900 -- 5400 \AA~ for velocity measurement,
  excluding the noisy region of $\lambda > 5500$ \AA~ due to some sky lines
  not completely eliminated even after sky subtraction.
Measuring errors of the radial velocity are typically $err(v)=35$ km s$^{-1}$.
For the objects with successfully measured velocity values,
  we measured the S/N values at $\lambda \sim 5000$ \AA,
  obtaining $1 <$S/N$< 75$.
The peak S/N values are 6 -- 10 for all these spectra, and 
  10 -- 20 for newly found, highest probability GCs.

\section{Cluster Search Method}\label{sec_search}

\ifemulate\figtwo\fi

We have used both photometric and spectroscopic information to select GCs in M31.
First, using photometric data, we investigated 
  various photometric parameters and morphological properties of the objects 
  in the CCD images.
Then we assigned spectral classes to bright objects,
  and used the radial velocities to determine the M31 membership of the objects
  with measured radial velocities. 
Finally, we performed the final classification 
  by careful visual inspection of the image
  of each object, after training our eyes 
  with images of the previously known GCs, stars, and galaxies in our own data.
Details of these steps are described below.

Before starting a survey of M31 GCs, 
  we tried to find the suitable parameter space to select M31 GC candidates 
  using the photometric data of the known M31 GCs.
We matched our photometric catalog of the objects
  with the previous catalogs of \citet{gal06} (Revised Bologna Catalog ver. 2.0
  [RBC2] -- their confirmed and candidate GCs),
  \citet{hux05}, and \citet{mac07}.
There are 861 objects (347 confirmed GCs and 514 GC candidates) common 
  between the previous catalogs and
  our catalog of photometry derived from our CCD images. 

\ifemulate\figthree\fi
\ifemulate\figfour\fi

Figure 2 shows the distributions of three SExtractor parameters 
  (stellarity, FWHM and ellipticity) based on the $R$ images, and 
  Figures 3 and 4 show the photometric diagrams of these objects.
In Figure 2 (a), (b), and (c),
  the crosses and the open solid histograms show
  the distributions of all 861 objects with good photometry,
  and the filled circles and hatched histograms show
  those of the confirmed GCs in common between this study and the papers above.
As stellarity of 1 corresponds to a point-source (star), and 
  a stellarity of 0 to a resolved object.
The distribution of stellarity in Figure 2 shows that
  most objects have stellarity of 1, few objects have stellarity between 0.1 and 0.8,
  and the rest have stellarity $\sim 0$.
Figure 2 (d) shows the histogram of the normalized FWHM, which is
  the measured FWHM divided by the seeing value of each image.

In Figure 2 several features are noted: 
(1) the distribution of the stellarity of the confirmed GCs shows 
  a strong peak around 1 with a broad tail extending to about 0.8, and
  a weak peak around 0. 
There are relatively much fewer GCs in the range between 0.1 and 0.8;
(2) the distribution of the ellipticity of the confirmed GCs shows a broad
peak around 0.1 with a tail extending to 0.5; and
(3) the distribution of the normalized FWHM of the confirmed GCs shows a strong peak
  around 1.4, which is significantly larger than that of the stars, 1.0.
Considering the pixel scale of our CCD chip,
  typical seeing of 2 pixels ($\approx 1.''4$), and
  the linear size of $1''$ at the distance of M31 ($\approx 3.78$ pc),
  it is expected that most of the GCs in M31 will appear point-source-like.
Even for these star-like GCs, Figure 2 (d) shows that the normalized FWHM
  is greater than 1.

Considering the features in Figure 2, we have set up two kinds of criteria 
  for the selection of GC candidates:
(1) criteria for ``all candidates'' are (a) all values of stellarity,
  (b) $1.15<$ FWHM/seeing $\lesssim 10$, and
  (c) ellipticity $<0.7$; and
(2) criteria for ``good candidates'' are 
  (a) stellarity of $0.8-1.0$,
  (b) $1.15<$ FWHM/seeing $\lesssim 5 $,
  and (c) ellipticity $<0.5$.
  ``Good candidates'' are candidates with higher probability 
  among ``all candidates.''
Figure 3 shows the $T_1 - (C-T_1)$ color-magnitude (CM) diagram (Fig 3 (a)), 
  the histograms of $(C-T_1)$ color (Fig 3 (b)) and $T_1$ magnitude (Fig 3 (c))
  of the 861 objects matched with the previous catalogs.
Figure 4 shows 
  the $(C-M)$--$(C-T_1)$ color-color (CC) diagram of the same objects.
Notable features in Figures 3 and 4 are that 
(1) the colors of the confirmed GCs are mostly in the range $1<(C-T_1)<2$, 
(2) the luminosity function (LF) of the confirmed GCs shows a peak at $T_1 \approx 17$ mag,
(3) the CM diagram shows a dominant vertical plume of GCs with $1<(C-T_1)<2$
  extending up to $T_1 \sim 13.5$, and
(4) the CC diagram shows a well-defined linear sequence of star clusters. 

In Figures 3 and 4 
  we also plotted the data for ``all candidates'' and ``good candidates''.
It is striking that the general properties of these candidates seen in these figures
  are very similar to those of the confirmed GCs, noting that no information of color 
  and magnitude was used for selecting these candidates.
This indicates that a significant fraction of these candidates may be GCs.

\ifemulate\figfive\fi

We selected the candidates in the best-seeing image
  among the $C, M,$ and $R$ images,
  which are predominantly the $R$-band images.
We used objects at least $\sim 1$ mag brighter than
  the limiting magnitude where the stellarity cannot be used to separate 
  stellar/non-stellar objects.
The typical limiting magnitude of the images is $T_1 = 22-23$ mag,
  varying from field to field due to the seeing and the crowding in the field.
We set the magnitude cut-off value for the candidate selection 
  1 -- 2 mag brighter than the limiting magnitude of the images, 
  depending on the seeing and the crowding in the images.
The magnitude cut-off value for the candidate selection ranges from 
  $T_1 \sim 17.5$ mag (for the center field) to 
  $\sim 19.5$ mag (for halo fields or good seeing), mostly $T_1 \sim 19$ mag.
Therefore our search is considered to be incomplete at $T_1>19$ mag for most fields
   (at $T_1 > 17.5$ mag for the central field).
Figure 5 shows an example of our application of the above criteria to one of 
  the KPNO fields (F56) with a seeing of $\approx 1.9$ pixels.
In this field, there are 4894 measured objects with good photometry.
Among these we selected 362 ``all GC candidates'' according to criterion (1),
 and 277 ``good GC candidates'' according to criterion (2). 

Finally we marked the ``all'' and ``good'' GC candidates selected above 
on the images, and 
visually inspected their images to finalize the GC candidates.
We checked contour maps and radial profiles of the objects
  as well as the images themselves.
In the contour map, we classified irregular, significantly elongated, asymmetric, 
  and loosely concentrated objects as galaxies, and 
  round, slightly elongated, strongly concentrated objects as star clusters.
Although some faint galaxies look round in the displayed images,
  in the contour maps their outer areas look irregular,
  while star clusters look very smooth and round.
Inspection of the contour maps was very
  efficient in selecting galaxies.
In the radial profile, 
the objects with FWHM larger than the seeing size
  are considered as GCs,
  those with FWHM similar to seeing size as stars, and
  those with a large excess in the wing as galaxies.
Checking color, position, and/or velocity information
  was also included. 
There are some confusing cases:
  compact elliptical galaxies versus GCs,
  compact GCs versus stars, and
  compact star clusters in {\sc{H~ii}} regions versus galaxies.
For these cases spectroscopic information was needed for classification.
In the outer areas close to the edge of each CCD image,
  the FWHMs get larger due to image degradation.
Therefore, we carefully compared potential targets in these areas
  with other nearby objects to see whether they are really extended.

\ifemulate\figsix\fi

Spectral information was also used for the classification of bright objects.
We have visually classified the flux-calibrated spectra into stars, star clusters, 
  and galaxies, comparing them with the template spectra of the spectral library of
  \citet{santos02}.
We used the continuum of the $4000-7000 \AA$ wavelength range as well as
  various spectral features:
  Balmer lines between 4000 and 5000 $\AA$ for early-type objects;
  absorption lines like {\sc{Ca~ii}} H and K, CH (G band), MgH$+$Mgb, and TiO for
  late-type objects; and emission lines for galaxies.
Figure 6 displays sample spectra for confirmed GCs,
  young star clusters, foreground stars (F, G, and K types), and three galaxies
  (M31, M32 and a background galaxy).

The radial velocities were used 
  as a strong constraint on the membership of the objects
  belonging to the Galaxy, M31, or the distant universe.
Most of the objects with radial velocities less than $-200$ km s$^{-1}$ 
  are probably M31 members,
while those from $-200$ to $+200$ km s$^{-1}$ could be M31 members 
  or Galactic foreground stars.
We considered all the objects with $v<-300$ km s$^{-1}$ to be M31 members.
The objects with $v>300$ km s$^{-1}$ were classified as background galaxies.
For objects with $-300$ km s$^{-1}$ $<v<+300$ km s$^{-1}$,
  we classified each object consulting its spectral class.

Combining both image and spectral inspections was 
  very efficient and accurate in classifying the objects.
For our fields F08--F42, we used both image inspection and 
  spectral inspection methods, 
  while for the fields without spectroscopic data 
  (F1--F7, F43--F52, and F56)
  we used only the image inspection method.

We have classified the final GCs/GC candidates into three classes according to 
  probability as follows: 
  (1) class 1, genuine GCs that were confirmed by either spectral types and 
              radial velocities or high resolution images (mostly $HST$ images);
  (2) class 2, probable clusters that are probably GCs from imaging data but 
              without spectral information; and 
  (3) class 3, possible clusters that are possibly GCs, 
  but may be other kinds of objects like background galaxies.

\section{Results} 
\subsection{The Catalog of New Globular Clusters in M31}
By applying the cluster search method described in the previous Section
  to the 53 KPNO fields of M31 covering a $\sim$ 3\arcdeg $\times$ 3\arcdeg~ area,
  we have found a total of 1164 GC candidates. 
Among these there are 605 new GC candidates found in this study
  and 559 previously known GCs in the catalogs of previous studies
  (e.g., \citet{hux05, gal06, mac07}).
Table 4 lists a summary of the numbers of the GCs and GC candidates
  for each class.
Among the new 605 GC candidates there are 113 genuine GCs (class 1),
  258 probable GCs (class 2) and 234 possible GCs (class 3).
Among the known GCs in previous studies we find 383 genuine GCs,
  109 probable GCs and 67 possible GCs.
In total there are 496 genuine GCs, 367 probable GCs and 301 possible GCs.

\ifemulate\figseven\fi
\ifemulate\figeight\fi
\ifemulate\fignine\fi

Tables $5-6$, $7-11$, and $12-15$
  present the lists of the GCs and GC candidates newly found 
  in this study for genuine GCs, probable GCs, 
  and possible GCs, respectively.
In these tables the columns give the running number (col. [1]), the coordinates 
  in right ascension and declination (J2000.0; cols. [2] and [3]), 
  $T_1$ magnitudes (col. [4]), $(M-T_1)$ and $(C-T_1)$ colors 
  (cols. [5] and [6]), and the radial velocities derived in this study (col. [7]).
The magnitudes and colors here are from the simple aperture photometry
  derived using SExtractor with an aperture of radius 5\arcsec.
Figures 7, 8, and 9 show the $R$-band mosaic images of genuine GCs (class 1) 
  with identifications in Tables $5-6$ labeled.

Tables $16-17$ and 18 
  present the lists of 111 stars and 21 galaxies, respectively,
  identified in the present study.
The columns are the same as those of Tables $5-6$, except the last column of Table 18,
  whithc gives the identification matched with the RBC2 of \citet{gal06}.
These lists of stars and galaxies were obtained from the spectral classification
  described in Section 3.
We found that 13 objects in these lists were already identified as galaxies
  or GC candidates without measured velocities in the RBC2 catalog.

\subsection{Properties of New Globular Clusters}

\ifemulate\figten\fi
\ifemulate\figeleven\fi
\ifemulate\figtwelve\fi

Although the detailed properties of the newly found GCs/GC candidates 
  as well as those of the whole M31 GC system including the previously known GCs 
  will be presented in separate papers,
  we show a few salient features of the newly found objects here.

Figure 10 shows the $T_1-(C-T_1)$ CM diagram (Fig. 10 (a)),
  $(C-T_1)$ color distributions (Fig. 10 (b)), and
  $T_1$ LF (Fig. 10 (c))
  of the newly found GCs/GC candidates in 
  Tables $5-6$, $7-11$, and $12-15$
Figure 11 shows the $(C-M)-(C-T_1)$ CC diagram of the same objects.
Most of the class 1 GCs have $T_1$ magnitudes of $17.5 - 19.5$ mag,
  which would be $V \approx 17.9 - 19.9$ mag assuming $(C-T_1) \approx 1.5$
  (see below) and using the
  \citet{gei96}'s transformation coefficients
  between $UBVRI$ photometry and $CT_1$ photometry.
Figure 12 shows the direct comparison of the confirmed, previously known GCs and
  newly found, class 1 GCs in the $T_1-(C-T_1)$ CM diagram, and
  it is noted that 
  most of the class 1 GCs newly found in this study are fainter than 
  most of the confirmed, previously known GCs. 
The brightest object among the newly found class 1 GCs has 
  $T_1 \sim 14.6$ ($V \sim 15.0$) mag and
  the faintest one has $T_1 \sim 19.9$ ($V \sim 20.1$) mag.
We checked the color distributions of the previously known GCs and
  newly found GCs, confirming that they are quite similar.

Figure 10 (b) shows that the $(C-T_1)$ color distributions of all three classes 
  encompass the color range of $1 < (C-T_1) < 2$ 
  [$0.6 \lesssim (B-V) \lesssim 1.0$; \citet{gei96}],
  in which most of the class 1 objects reside.
There are a rather large number of very red objects with $2 < (C-T_1) < 2.5$
  [$(B-V) \approx 1.0 - 1.3$], 
  which could be reddened GCs or intrinsically red clusters.

\ifemulate\figthirteen\fi

Figure 13 shows the spatial distribution and histograms
  of the newly found GCs and GC candidates.
Open circles and filled histograms are for class 1 GCs in this study,
  crosses are for class 2 GCs in this study, and
  dots and open solid histograms are for GCs in
  the catalogs of \citet[their class 1]{gal06},
  \citet{hux05}, and \citet{mac07}.
Most of the newly found, class 1 GCs are located in the disk area of M31.
Higher spatial resolution imaging and spectroscopy, and possibly
  in the near-infrared wavelength band, would be needed
  to search for GCs in the central region of M31
  where we missed many faint GCs, as seen in Figure 13 ($top$).

\section{Summary}

We have presented the results of a new systematic wide field CCD survey of M31 GCs.
  Using Washington $CMT_1$ CCD images
  obtained at the KPNO 0.9 m telescope
  and spectra obtained using the WIYN 3.5 m telescope and Hydra 
  multifiber bench spectrograph,
  we have investigated the photometric and morphological parameters of the objects,
  visually checked their images,
  and obtained their spectra and radial velocities.
Finally, we have found 1164 GCs and GC candidates,
  of which 559 are previously known GCs
  and 605 are newly found GC candidates.
Among the new objects there are 113 genuine GCs (class 1),
  258 probable GCs (class 2), and 234 possible GCs (class 3).
Among the previously known objects there are 383 genuine GCs,
  109 probable GCs and 67 possible GCs.
In total there are 496 genuine GCs, 367 probable GCs and 301 possible GCs.

The magnitudes and colors of 
  most of the newly found class 1 objects are $17.5 < T_1 < 19.5$ mag
  and $ 1 < (C-T_1)< 2$.
The faintest part of the M31 GC LF
  is mostly filled with these new GC candidates,
  although the intrinsically very faint GCs like 
  AM 4, Palomar 1, E 3, and Palomar 13 
  in the Galaxy (see, e.g., \citet{vdB04, sar07}) may remain
  to be detected.

\vskip 3cm
We would like to thank the anonymous referee 
  for providing prompt and thoughtful comments that helped improve 
  the original manuscript.
The authors are grateful to the staff members of the KPNO
  for their warm support during our observations and data reduction.
The WIYN Observatory is a joint facility of the University of 
  Wisconsin-Madison, Indiana University, Yale University, and 
  the National Optical Astronomy Observatory. 
M. G. L. was supported in part by a Korean Research Foundation grant
  (KRF-2000-DP0450) and ABRL (R14-2002-058-01000-0).
D. G. gratefully acknowledges support from Chilean Centro de 
  Astrof{\'i}sica FONDAP grant 15010003.
A. S. was supported by NSF CAREER grant AST 00-94048.


\begin{table}
\scriptsize
\begin{center}
\caption{A List of Previous M31 GC Searches}
\begin{tabular}{ccccc}
\hline\hline
Reference & Plate vs. CCD & CCD FOV & N(GC)& Comments \\
\hline
\citet{hub32} & plate & \nodata &  140  & M31's disk only \\
\citet{sey45} & plate & \nodata &  101  & W. Baade's discovery, no coordinates \\
\citet{vet62} & plate & \nodata & (241)\tablenotemark{a} & coordinates, $V$ magnitudes
  and colors \\
\citet{baa64} & plate & \nodata &   30  & \\
\citet{may53} & plate & \nodata &    4  & outer parts of M31 \\
\citet{all76} & plate & \nodata &    5  & nuclear region of M31 \\
\citet{sar77} & plate & \nodata &  355  & KPNO 4 m \\
\citet{cra85} & spectra plates & \nodata & 109 & CFHT 3.6 m, A catalog of total 509 GCs \\
\citet{bat87} & plate & \nodata &  353\tablenotemark{b}  & Bologna 1.52 m \\ 
\citet{wir85} & video camera & $2' \times 2'$ & $\sim 50$ & bulge region \\
\citet{aur92} & CCD   & $1'.7 \times 2'.7$ & 16\tablenotemark{c} & 
  $7'.7 \times 7'.7$ field centered on M31 \\
\citet{bat93} & CCD   & $320 \times 512$ pixel$^2$ & 
   4\tablenotemark{d} & $R < 5.'5$ ($R \lesssim 1$ kpc) \\ 
\citet{moc98} & CCD   & $11' \times 11'$ & 4\tablenotemark{e} &
   four fields in the M31's disk \\
\citet{bar00} & CCD   & $22' \times 22'$ & (435)\tablenotemark{f} & A new catalog of ``best'' photometry \\
\citet{bar01} & HST/WFPC2 & $\sim 2'.7 \times 2'.7$ & 32 & 157 images \\
\citet{per02} & spectroscopy & \nodata & (288)\tablenotemark{g} & WHT 4.2 m + WYFFOS \\
\citet{gal04} & 2MASS/NICMOS3 & all sky & (693)\tablenotemark{h} &
   Revised Bologna Catalogue of 1035 objects \\ 
\citet{hux05} & CCD   & $\approx 0.29$deg$^2$ & 3 & INT 2.5 m + WFC \\
\citet{gal06} & spectroscopy & \nodata & (42)\tablenotemark{i} & Revised Bologna Catalogue V2.0\\
\citet{mac06,mac07} & HST/ACS & $\sim 3'.4 \times 3'.4$ & 14 & HST Program GO 10394
   (Cycle 13) \\
This study    & CCD & $23.'2 \times 23.'2$ & 605\tablenotemark{j} & mapping 
$\sim 3\arcdeg \times 3\arcdeg$ field centered on M31 \\
\hline
\end{tabular}
\tablenotetext{a}{Not new discoveries, but studies
  on the objects found by \citet{hub32} and \citet{sey45}.}
\tablenotetext{b}{With 254 class A, 99 class B, 152 class C, and 218 class D objects,
  where
  class A objects are very high-confidence objects,
  class B objects are high confidence objects,
  class C objects are plausible candidates, and
  class D miscellaneous non-stellar objects with an expected percentage
  of actual clusters of the order of a few percent.}
\tablenotetext{c}{With 12 reliable, 4 possible.}
\tablenotetext{d}{With 3 class A, 1 class B, 20 class C, and 20 class D objects.}
\tablenotetext{e}{With 2 class A, 2 class B, 28 class C, and 36 class D objects.}
\tablenotetext{f}{Observations of 13 fields centered on M31, but no GC search.
  Of the 435 objects, 268 have optical photometry in four or more filters,
  224 have near-infrared photometry, 200 have radial velocities, and
  188 have spectroscopic metallicities.}
\tablenotetext{g}{Spectroscopy for 288 previously known objects.
  They presented a spectroscopic database of 321 velocities and 301 metallicities.}
\tablenotetext{h}{2MASS near-infrared photometry for 693 known and candidate GCs.
  Of 1035 objects, 337 are confirmed GCs,
  688 are GC candidates, and 10 are objects with controversial classification.}
\tablenotetext{i}{Confirmed GC nature from spectroscopy of 76 candidates.}
\tablenotetext{j}{With 113 class 1, 258 class 2, and 234 class 3, where
   classes 1, 2, and 3 are similar to classes A, B, and C, respectively,
   in \citet{bat87}.}
  \end{center}
\end{table}

\begin{table} 
\footnotesize
\begin{center}
\caption{Summary of KPNO 0.9 m Photometric Observations\label{phobslog}}
{\tiny  
\vspace{0.3cm}
\setlength{\tabcolsep}{1.2mm}
\begin{tabular}{l l l l l l}
\hline\hline
Night & Obs. Date (UT) & Field & Weather \\
  (1) &     (2)        & (3)   & (4) \\
\hline
 N1 & 1996 October 14 & 23($C$) 26                       & cloudy \\
 N2 & 1996 October 15 & 16 17 18(long)$^{\rm a}$ 19 20 23($M,T_1$) 24 25(long)$^{\rm b}$ 27
                                          & cloudy \\
 N3 & 1996 October 16 & {\bf 18(short)$^{\rm a}$ 25(short)$^{\rm b}$ 30 31 32 33 34 35} &
                                          {\bf semi-photometric} \\
 N4 & 1996 October 17 & {\bf 28 36} & {\bf semi-photometric} \\
 N5 & 1996 October 18 & {\bf 09 10 37 38 39 40 41} & {\bf photometric} \\
 N6 & 1996 October 19 & \nodata   & \nodata \\
 N7 & 1996 October 20 & \nodata   & \nodata \\
 N8 & 1996 October 21 & {\bf 11 12 13 14 15 29 42 43($M$)} & {\bf photometric} \\
 N9 & 1996 October 22 & {\bf 43($C,T_1$) 44 45$^{\rm c}$} & {\bf photometric} \\
N10 & 1996 October 23 & (45$^{\rm c}$) 46 47 48 49 50  & non-photometric \\
N11 & 1996 October 24 & {\bf 01 02 03 04 05}    & {\bf semi-photometric} \\
N12 & 1996 October 25 & 06                 & non-photometric \\
Oct98 & 1998 October 19 & {\bf 07 08 21 22 51 52 56} & {\bf photometric} \\
\hline
\end{tabular}
} 
\end{center}
\tablecomments{The fields with standard transformation data (photometric and
   semi-photometric nights) are represented with bold letters. \\
$^{\rm a}$ For the field of F18, which is just the northern field of F25, we took 
   two sets of data; long exposures (1500s in $C$, 600s in $M$, and 600s in $T_1$)
   and short exposures (300s in $C$, 100s in $M$, and 100s $\times$ 2 in $T_1$). \\
$^{\rm b}$ For the field of F25, which includes the M31 central region, we took 
   of data; long exposures (1500s in $C$, 600s in $M$, and 300s $\times$ 2 in $T_1$)
   and short exposures (300s in $C$, 200s in $M$, and 200s $\times$ 2 in $T_1$). \\
$^{\rm c}$ Field 45 was observed both on N9 and N10 with the same exposure time setups.
   Even though the seeing of N10 for F45 was slightly better than that of N9
   ($1.''6$ versus $2.''0$),
   we primarily used the N9 data for the utilization
   of N9 standard transformation information.}
\end{table} 

\begin{table} 
\begin{center}
\caption{Summary of WIYN 3.5 m Spectroscopic Observations\label{spobslog}}
\scriptsize  
\begin{tabular}{lllll}
\hline\hline
               & Hydra         & N(observed)\tablenotemark{a} & &  \\
Obs. Date (UT) & Configuration & (=new\tablenotemark{b} + old\tablenotemark{c}) 
  & N(vel)\tablenotemark{d} & Exposure Time \\
\hline
2000 September 7 & 1 & \phantom{0}68 (=\phantom{0}66+\phantom{0}2) & 38 & 1800 s $\times$ 4 = 7200 s (2$^h$) \\
2000 September 7 & 2 & \phantom{0}73 (=\phantom{0}56+17)           & 66 & 1800 s $\times$ 3 + 900 s = 6300 s (1$^h$ 45$^m$) \\
2000 September 8 & 3 & \phantom{0}64 (=\phantom{0}64+\phantom{0}0) & 47 & 1800 s $\times$ 4 = 7200 s (2$^h$) \\
2000 September 8 & 4 & \phantom{0}62 (=\phantom{0}59+\phantom{0}3) & 34 & 1800 s $\times$ 3 = 5400 s (1$^h$ 30$^m$) \\
2000 September 8 & 5 & \phantom{0}71 (=\phantom{0}59+12)           & 56 & 1800 s $\times$ 4 = 7200 s (2$^h$) \\
2000 September 9 & 6 & \phantom{0}75 (=\phantom{0}74+\phantom{0}1) & 36 & 1800 s $\times$ 4 = 7200 s (2$^h$) \\
2000 September 9 & B & \phantom{0}70 (=\phantom{0}67+\phantom{0}3) & 56 & 1800 s $\times$ 4 = 7200 s (2$^h$) \\
2000 September 9 & E & \phantom{0}70 (=\phantom{0}60+10)           & 57 & 1800 s $\times$ 3 + 1041 s = 6441 s (1$^h$ 47$^m$ 21$^s$) \\
                 & sub-total & 552 (=504+48) & 390 & \\
\hline
2001 November 2 & 11 & \phantom{0}69 (=\phantom{0}54+15) & 35 &     2400 s $\times$ 3 = 7200 s (2$^h$) \\
2001 November 2 & 12 & \phantom{0}79 (=\phantom{0}45+34) & 60 &     2400 s $\times$ 3 = 7200 s (2$^h$) \\
2001 November 2 & 17 & \phantom{0}47 (=\phantom{0}38+\phantom{0}9) & 20 & 2400 s $\times$ 1 = 2400 s (40$^m$) \\
                & sub-total  & 195 (=137+58) & 115 & \\
\hline
                & Total      & 748 (=642+106) & 505 & \\
\hline
\end{tabular}
\tablenotetext{a}{Number of all observed objects.}
\tablenotetext{b}{GC candidates found from the photometric criteria in this study.}
\tablenotetext{c}{Previously known GCs.} 
\tablenotetext{d}{Number of objects for which successful velocities were measured.}
  \end{center}
\end{table}

\begin{table} 
\begin{center}
\caption{Number of Globular Clusters Found in This Study\label{GCnum}}
\scriptsize
\begin{tabular}{cccc}
\hline\hline
Class & Previously found by others\tablenotemark{a} & New GCs & Sum \\
\hline
1  & 383  & 113 & 496 \\
2  & 109  & 258 & 367 \\
3  &  67  & 234 & 301 \\
\hline
Sum & 559 & 605 & 1164 \\
\hline
\end{tabular}
\tablenotetext{a}{in \citet{hux05, gal06, mac07} and references there in.}
\end{center}
\end{table}

\clearpage
\LongTables 
\begin{deluxetable}{crrrrrr}
\tabletypesize{\scriptsize} 
\tablecaption{A List of New Globular Clusters (class 1)\label{tab5}}
\tablewidth{-3pt}
\tablehead{
\colhead{}& \colhead{R.A.\tablenotemark{a}} &
\colhead{Decl.\tablenotemark{b}} &
\colhead{}& \colhead{}& \colhead{}& \colhead{$v$} \\
  \colhead{ID} & \colhead{(J2000.0)} &
  \colhead{(J2000.0)} &
  \colhead{$T_1$} & \colhead{$(M-T_1)$} & 
  \colhead{$(C-T_1)$} & \colhead{(km s$^{-1}$)} \\
\colhead{(1)}& \colhead{(2)}&
\colhead{(3)}&
\colhead{(4)}& \colhead{(5)}& \colhead{(6)}& \colhead{(7)}
} 
\startdata  
  1 & 00:36:33.54 & +40:39:44.3& $17.251\pm 0.007$ & $ 0.650\pm 0.011$ & $ 1.542\pm 0.015$ & $ -239.5\pm   46.7 $\\
  2 & 00:36:34.99 & +41:01:08.0& $18.805\pm 0.030$ & $ 0.677\pm 0.047$ & $ 1.407\pm 0.060$ & $ -402.7\pm   92.5 $\\
  3 & 00:37:10.67 & +41:41:17.4& $17.834\pm 0.011$ & $ 0.778\pm 0.019$ & $ 1.787\pm 0.030$ & $ -315.9\pm   34.5 $\\
  4 & 00:38:01.31 & +42:04:06.0& $18.311\pm 0.016$ & $ 0.708\pm 0.027$ & $ 1.565\pm 0.040$ & $  -27.4\pm   32.8 $\\
  5 & 00:38:59.15 & +41:40:27.4& $18.848\pm 0.031$ & $ 0.877\pm 0.077$ & $ 1.842\pm 0.083$ & $ -405.2\pm   43.1 $\\
  6 & 00:39:39.69 & +40:33:56.4& $18.811\pm 0.045$ & $ 0.648\pm 0.062$ & $ 1.456\pm 0.078$ & $ -319.9\pm   53.5 $\\
  7 & 00:39:44.03 & +40:51:18.4& $17.630\pm 0.017$ & $ 0.204\pm 0.020$ & $-0.068\pm 0.018$ & $ -390.4\pm   86.6 $\\
  8 & 00:40:00.39 & +40:43:53.5& $18.581\pm 0.038$ & $ 0.769\pm 0.057$ & $ 1.628\pm 0.070$ & $ -108.9\pm   27.9 $\\
\enddata 
\tablecomments{Table 5 is published in its entirety in the electronic edition
  of the {\it Astronomical Journal}.  A portion is shown here for guidance
  regarding its form and content.}
\tablenotetext{a}{Right ascension in units of hours, minutes, and seconds.}
\tablenotetext{b}{Declination in units of degrees, arcminutes, and arcseconds.}
\end{deluxetable}

\clearpage
\LongTables 
\begin{deluxetable}{crrrrrr}
\tabletypesize{\scriptsize}  
\tablecaption{A List of New Probable Globular Clusters (class 2)\label{tab6}}
\tablewidth{-3pt}
\tablehead{ 
\colhead{}& \colhead{R.A.\tablenotemark{a}} &
\colhead{Decl.\tablenotemark{b}} &
\colhead{}& \colhead{}& \colhead{}& \colhead{$v$} \\
  \colhead{ID} & \colhead{(J2000.0)} &
  \colhead{(J2000.0)} &
  \colhead{$T_1$} & \colhead{$(M-T_1)$} &
  \colhead{$(C-T_1)$} & \colhead{(km s$^{-1}$)} \\
\colhead{(1)}& \colhead{(2)}&
\colhead{(3)}& 
\colhead{(4)}& \colhead{(5)}& \colhead{(6)}& \colhead{(7)}
}
\startdata
  1 & 00:33:23.07 & +40:04:40.7& $17.784\pm 0.015$ & $ 0.896\pm 0.036$ & $ 2.223\pm 0.102$ & \nodata\phn\phn\phn\phn\\
  2 & 00:33:32.20 & +39:51:32.8& $17.534\pm 0.010$ & $ 0.505\pm 0.014$ & $ 1.009\pm 0.014$ & \nodata\phn\phn\phn\phn\\
  3 & 00:33:37.03 & +39:40:59.0& $16.283\pm 0.004$ & $ 0.794\pm 0.006$ & $ 1.570\pm 0.007$ & \nodata\phn\phn\phn\phn\\
  4 & 00:34:34.20 & +40:02:49.4& $18.777\pm 0.038$ & $ 1.145\pm 0.106$ & $ 2.277\pm 0.266$ & \nodata\phn\phn\phn\phn\\
  5 & 00:35:21.20 & +41:55:32.6& $17.957\pm 0.013$ & $ 1.031\pm 0.025$ & $ 2.646\pm 0.065$ & \nodata\phn\phn\phn\phn\\
  6 & 00:35:34.24 & +41:11:53.0& $18.409\pm 0.018$ & $ 0.883\pm 0.033$ & $ 1.826\pm 0.050$ & \nodata\phn\phn\phn\phn\\
  7 & 00:35:45.26 & +39:39:21.3& $18.334\pm 0.020$ & $ 0.943\pm 0.036$ & $ 2.435\pm 0.082$ & \nodata\phn\phn\phn\phn\\
  8 & 00:35:58.15 & +39:37:35.5& $17.527\pm 0.010$ & $ 0.963\pm 0.018$ & $ 1.938\pm 0.028$ & \nodata\phn\phn\phn\phn\\
\enddata
\tablecomments{Table 6 is published in its entirety in the electronic edition
  of the {\it Astronomical Journal}.  A portion is shown here for guidance
  regarding its form and content.}
\tablenotetext{a}{Right ascension in units of hours, minutes, and seconds.}
\tablenotetext{b}{Declination in units of degrees, arcminutes, and arcseconds.}
\end{deluxetable}

\clearpage
\LongTables 
\begin{deluxetable}{crrrrrr}
\tabletypesize{\scriptsize}  
\tablecaption{A List of New Possible Globular Clusters (class 3)\label{tab7}}
\tablewidth{-3pt}
\tablehead{ 
\colhead{}& \colhead{R.A.\tablenotemark{a}} &
\colhead{Decl.\tablenotemark{b}} &
\colhead{}& \colhead{}& \colhead{}& \colhead{$v$} \\
  \colhead{ID} & \colhead{(J2000.0)} &
  \colhead{(J2000.0)} &
  \colhead{$T_1$} & \colhead{$(M-T_1)$} &
  \colhead{$(C-T_1)$} & \colhead{(km s$^{-1}$)} \\
\colhead{(1)}& \colhead{(2)}&
\colhead{(3)}& 
\colhead{(4)}& \colhead{(5)}& \colhead{(6)}& \colhead{(7)}
}
\startdata
  1 & 00:33:13.08 & +40:05:26.0& $18.219\pm 0.023$ & $ 0.990\pm 0.057$ & $ 1.998\pm 0.127$ & \nodata\phn\phn\phn\phn\\
  2 & 00:33:15.82 & +40:00:24.7& $18.018\pm 0.019$ & $ 0.968\pm 0.046$ & $ 2.502\pm 0.159$ & \nodata\phn\phn\phn\phn\\
  3 & 00:33:34.43 & +40:11:06.5& $17.231\pm 0.009$ & $ 0.897\pm 0.022$ & $ 2.013\pm 0.052$ & \nodata\phn\phn\phn\phn\\
  4 & 00:33:34.96 & +40:08:16.3& $18.523\pm 0.030$ & $ 1.166\pm 0.087$ & $ 1.233\pm 0.093$ & \nodata\phn\phn\phn\phn\\
  5 & 00:33:38.04 & +39:35:35.7& $18.067\pm 0.015$ & $ 1.182\pm 0.032$ & $ 2.798\pm 0.069$ & \nodata\phn\phn\phn\phn\\
  6 & 00:33:46.11 & +39:48:36.6& $18.163\pm 0.017$ & $ 0.990\pm 0.031$ & $ 2.217\pm 0.049$ & \nodata\phn\phn\phn\phn\\
  7 & 00:33:54.63 & +39:34:36.6& $17.644\pm 0.010$ & $ 1.311\pm 0.024$ & $ 2.909\pm 0.052$ & \nodata\phn\phn\phn\phn\\
  8 & 00:34:11.50 & +40:07:58.0& $18.381\pm 0.026$ & $ 0.974\pm 0.065$ & $ 2.007\pm 0.148$ & \nodata\phn\phn\phn\phn\\
\enddata
\tablecomments{Table 7 is published in its entirety in the electronic edition
  of the {\it Astronomical Journal}.  A portion is shown here for guidance
  regarding its form and content.}
\tablenotetext{a}{Right ascension in units of hours, minutes, and seconds.}
\tablenotetext{b}{Declination in units of degrees, arcminutes, and arcseconds.}
\end{deluxetable}  

\clearpage
\LongTables 
\begin{deluxetable}{crrrrrr}
\tabletypesize{\scriptsize}  
\tablecaption{A List of Stars Identified in this Study\label{tab8}}
\tablewidth{-3pt}
\tablehead{
\colhead{}& \colhead{R.A.\tablenotemark{a}} &
\colhead{Decl.\tablenotemark{b}} &
\colhead{}& \colhead{}& \colhead{}& \colhead{$v$} \\
  \colhead{ID} & \colhead{(J2000.0)} &
  \colhead{(J2000.0)} &
  \colhead{$T_1$} & \colhead{$(M-T_1)$} &
  \colhead{$(C-T_1)$} & \colhead{(km s$^{-1}$)} \\
\colhead{(1)}& \colhead{(2)}&
\colhead{(3)}&
\colhead{(4)}& \colhead{(5)}& \colhead{(6)}& \colhead{(7)}
}
\startdata
  1 & 00:35:10.78 & +41:44:03.1 & $ 16.340\pm 0.003$ & $  0.711\pm 0.005$ & $  1.701\pm 0.008$ & $   -15.9\pm   37.2$ \\
  2 & 00:35:17.16 & +41:22:13.0 & $ 16.014\pm 0.003$ & $  0.707\pm 0.004$ & $  1.669\pm 0.006$ & $   -54.5\pm   28.2$ \\
  3 & 00:35:20.08 & +41:21:41.7 & $ 15.935\pm 0.002$ & $  0.590\pm 0.004$ & $  1.341\pm 0.005$ & $    -1.0\pm   27.0$ \\
  4 & 00:35:25.19 & +42:06:51.3 & $ 16.863\pm 0.006$ & $  0.627\pm 0.009$ & $  1.344\pm 0.011$ & $  -102.1\pm   38.0$ \\
  5 & 00:35:26.09 & +41:44:42.0 & $ 15.639\pm 0.002$ & $  0.618\pm 0.003$ & $  1.421\pm 0.004$ & $   -66.8\pm   26.1$ \\
  6 & 00:35:33.58 & +42:07:06.8 & $ 14.486\pm 0.001$ & $  0.728\pm 0.002$ & $  1.730\pm 0.003$ & $   -38.7\pm   24.0$ \\
  7 & 00:35:37.69 & +42:07:03.5 & $ 15.766\pm 0.002$ & $  0.732\pm 0.004$ & $  1.707\pm 0.006$ & $    40.9\pm   43.1$ \\
  8 & 00:35:45.46 & +41:45:14.0 & $ 15.431\pm 0.002$ & $  0.581\pm 0.003$ & $  1.366\pm 0.004$ & $   -33.3\pm   24.7$ \\
\enddata
\tablecomments{Table 8 is published in its entirety in the electronic edition
  of the {\it Astronomical Journal}.  A portion is shown here for guidance
  regarding its form and content.}
\tablenotetext{a}{Right ascension in units of hours, minutes, and seconds.}
\tablenotetext{b}{Declination in units of degrees, arcminutes, and arcseconds.}
\end{deluxetable}  

\begin{deluxetable}{crrrrrrc}
\tabletypesize{\scriptsize}  
\tablecaption{A List of Galaxies Identified in this Study\label{tab9}}
\tablewidth{-3pt}
\tablehead{
\colhead{}& \colhead{R.A.\tablenotemark{a}} &
\colhead{Decl.\tablenotemark{b}} &
\colhead{}& \colhead{}& \colhead{}& \colhead{$v$}& \\
  \colhead{ID} & \colhead{(J2000.0)} &
  \colhead{(J2000.0)} &
  \colhead{$T_1$} & \colhead{$(M-T_1)$} &
  \colhead{$(C-T_1)$} & \colhead{(km s$^{-1}$)}& \colhead{RBC2 ID\tablenotemark{c}} \\
\colhead{(1)}& \colhead{(2)}&
\colhead{(3)}&
\colhead{(4)}& \colhead{(5)}& \colhead{(6)}& \colhead{(7)}& \colhead{(8)}
}
\startdata
 1 & 00:36:32.94 & +40:24:52.7 & $16.408\pm 0.004 $& $  0.788\pm 0.006 $& $  1.849\pm 0.009 $& $ 10779.3\pm   51.1 $&    B294  \\
 2 & 00:40:15.67 & +41:48:30.4 & $17.252\pm 0.008 $& $  0.841\pm 0.014 $& $  1.835\pm 0.022 $& $ 53856.7\pm   68.2 $&    B320  \\
 3 & 00:40:27.23 & +41:29:09.1 & $17.352\pm 0.014 $& $  0.801\pm 0.018 $& $  1.705\pm 0.023 $& $ 41878.8\pm   69.1 $&    B007  \\
 4 & 00:40:55.18 & +41:57:28.7 & $17.695\pm 0.011 $& $  0.881\pm 0.021 $& $  2.174\pm 0.041 $& $ 35348.9\pm   74.1 $&  \nodata \\
 5 & 00:41:00.13 & +41:55:39.8 & $16.972\pm 0.006 $& $  0.951\pm 0.012 $& $  2.260\pm 0.023 $& $ 28551.8\pm   65.7 $&  \nodata \\
 6 & 00:42:49.35 & +41:59:06.6 & $18.392\pm 0.032 $& $  0.732\pm 0.056 $& $  2.059\pm 0.138 $& $ 35984.0\pm  100.3 $&  \nodata \\
 7 & 00:43:14.56 & +40:22:58.9 & $18.767\pm 0.028 $& $  1.004\pm 0.052 $& $  2.532\pm 0.116 $& $ 38863.2\pm   89.3 $&  \nodata \\
 8 & 00:43:15.67 & +42:01:47.9 & $17.738\pm 0.018 $& $  0.759\pm 0.031 $& $  2.045\pm 0.075 $& $ 22246.6\pm   58.1 $&  \nodata \\
\enddata
\tablenotetext{a}{Right ascension in units of hours, minutes, and seconds.}
\tablenotetext{b}{Declination in units of degrees, arcminutes, and arcseconds.}
\tablenotetext{c}{Identification in the Revised Bologna Catalogue V2.0 of
  \citet{gal06}} 
\end{deluxetable}  


\begin{thebibliography}{} 
\bibitem[Alloin, Pelat, \& Bijaoui (1976)]{all76} 
   Alloin, D., Pelat, D., \& Bijaoui, A. 1976,
   \aap, 50, 127 (Erratum 1977, \aap, 54, 321)
\bibitem[Auri{\`e}re, Coupinot, \& Hecquet (1992)]{aur92} 
   Auri{\`e}re, M., Coupinot, G., \& Hecquet, J. 1992,
   \aap, 256, 95    
\bibitem[Baade \& Arp (1964)]{baa64} 
   Baade, W., \& Arp, H. C. 1964, \apj, 139, 1027
\bibitem[Barmby et al. (2000)]{bar00} 
   Barmby, P., Huchra, J. P., Brodie, J. P., Forbes, D. A., Schroder, L. L.,
   \& Grillmair, C. J. 2000, \aj, 119, 727 
\bibitem[Barmby \& Huchra (2001)]{bar01} 
   Barmby, P., \& Huchra, J. P. 2001, \aj, 122, 2458
\bibitem[Battistini et al. (1987)]{bat87} 
   Battistini, P., B\`onoli, F., Braccesi, A., Federici, L., Fusi Pecci, F.,
   Marano, B., \& B\"orngen, F. 1987, \aaps, 67, 447
\bibitem[Battistini et al. (1993)]{bat93} 
   Battistini, P., B\`onoli, F., Casavecchia, M., Ciotti, L., Federici, L.,
   \& Fusi Pecci, F. 1993, \aap, 272, 77
\bibitem[Bertin \& Arnouts (1996)]{ber96} Bertin, E., \& Arnouts, S. 1996, \aaps, 117, 393
\bibitem[Crampton et al. (1985)]{cra85} 
   Crampton, D., Cowley, A. P., Schade, D., \& Chayer, P. 1985,
      \apj, 288, 494 
\bibitem[De Angeli et al. (2005)]{dea05} De Angeli, F., Piotto, G., Cassisi, S., 
   Busso, G., Recio-Blanco, A., Salaris, M., Aparicio, A., \& Rosenberg, A.
   2005, \aj, 130, 116 
\bibitem[Galleti et al. (2004)]{gal04} Galleti, S., Federici, L., Bellazzini, M.,
   Fusi Pecci, F., \& Macrina, S. 2004, \aap, 416, 917
\bibitem[Galleti et al. (2006)]{gal06} Galleti, S., Federici, L., Bellazzini, M.,
   Buzzoni, A., \& Fusi Pecci, F. 2006, \aap, 456, 985
\bibitem[Geisler (1996)]{gei96} Geisler, D. 1996, \aj, 111, 480
\bibitem[Hubble (1932)]{hub32} 
   Hubble, E. P. 1932, \apj, 76, 44
\bibitem[Huchra, Brodie, \& Kent (1991)]{huc91} 
   Huchra, J. P., Brodie, J. P., \& Kent, S. M. 1991,
   \apj, 370, 495 
\bibitem[Huxor \etal (2005)]{hux05}Huxor, A. P., Tanvir, N. R., Irwin, M. J.,
  Ibata, R., Collett, J. L., Ferguson, A. M. N., Bridges, T., \& Lewis, G. F.
  2005, \mnras, 360, 1007 
\bibitem[Karachentsev et al. (2004)]{kara04}
  Karachentsev, I. D., Karachentseva, V. E., Huchtmeier, W. K., \& Makarov, D. I.
  2004, AJ, 127, 2031
\bibitem[Kim \etal (2002)]{kim02} Kim, S. C., Lee, M. G., Geisler, D., 
  Seguel, J., Sarajedini, A., \& Harris, W. E. 2002, 
  in IAU Symp. 207, Extragalactic Star Clusters, ed. D. Geisler, E. K. Grebel,
  \& D. Minniti, 143 
\bibitem[Kodaira \etal (2004)]{kod04}Kodaira, K., Vansevicius, V., Bridzius, A., 
  Komiyama, Y., Miyazaki, S., Stonkute, R., Sabelviciute, I., \& Narbutis, D. 
  2004, \pasj, 56, 1025
\bibitem[Lee \etal (2002)]{lee02} Lee, M. G., Kim, S. C., Geisler, D., 
  Seguel, J., Sarajedini, A., \& Harris, W. E. 2002, 
  in IAU Symp. 207, Extragalactic Star Clusters, ed. 
  D. Geisler, E. K. Grebel, \& D. Minniti, 46 
\bibitem[Mackey \etal (2006)]{mac06}Mackey, A. D., Huxor, A., Ferguson, A. M. N.,
  Tanvir, N. R., Irwin, M.,  Ibata, R., Bridges, T., Johnson, R. A., \&
  Lewis, G. 2006, ApJ, 653, L105 
\bibitem[Mackey \etal (2007)]{mac07}Mackey, A. D., Huxor, A., Ferguson, A. M. N.,
  Tanvir, N. R., Irwin, M.,  Ibata, R., Bridges, T., Johnson, R. A., \&
  Lewis, G. 2007, ApJ, 655, L85 
\bibitem[Martin \etal (2006)]{mar06}Martin, N. F., Ibata, R. A., Irwin, M. J.,
  Chapman, S., Lewis, G. F., Ferguson, A. M. N., Tanvir, N., \& McConnachie, A. W.
  2006, \mnras, 371, 1983
\bibitem[Mayall \& Eggen (1953)]{may53} 
   Mayall, N. U., \& Eggen, O. J. 1953, \pasp, 65, 24
\bibitem[Mochejska et al. (1998)]{moc98}
   Mochejska, B. J., Kaluzny, J., Krockenberger, M., Sasselov, D. D.,
   \& Stanek, K. Z. 1998, Acta Astronomica, 48, 455
\bibitem[Morrison et al. (2003)]{morrison03}
  Morrison, H. L., Harding, P., Hurley-Keller, D., \& Jacoby, G.
  2003, ApJ, 596, L183
\bibitem[Perrett et al. (2002)]{per02} 
   Perrett, K. M., Bridges, T. J., Hanes, D. A., Irwin, M. J., Brodie, J. P.,
   Carter, D., Huchra, J. P., \& Watson, F. G. 2002, \aj, 123, 2490 
\bibitem[Racine (1991)]{racine91}
   Racine, R. 1991, AJ, 101, 865
\bibitem[Salaris \& Weiss (2002)]{sal02}
   Salaris, M., \& Weiss, A. 2002, \aap, 388, 492
\bibitem[Santos et al. (2002)]{santos02}
   Santos, J. F. C., Jr., Alloin, D., Bica, E., \& Bonatto, C. 2002,
   in Extragalactic Star Clusters, IAU Symp. 207, eds.
   D. Geisler, E. K. Grebel, \& D. Minniti, 727
\bibitem[Sarajedini et al. (2007)]{sar07} Sarajedini, A., Bedin, L. R.,
  Chaboyer, B., Dotter, A., Siegel, M., Anderson, J., Aparicio, A., 
  King, I., Majewski, S., Mar{\'i}n-Franch, A., Piotto, G., Reid, I. N.,
  \& Rosenberg, A. 2007, AJ, accepted (astro-ph/0612598)
\bibitem[Sargent et al. (1977)]{sar77} 
   Sargent, W. L. W., Kowal, C. T., Hartwick, F. D. A., \& van den Bergh, S.
   1977, \aj, 82, 947
\bibitem[Seguel et al. (2002)]{seg02} Seguel, J., Geisler, D., Lee, M. G., 
   Kim, S. C., Sarajedini, A., \& Harris, W. E. 2002, in IAU Symp. 207, 
   Extragalactic Star Clusters, ed. D. Geisler, E. K. Grebel, \& D. Minniti, 146
\bibitem[Seyfert \& Nassau (1945)]{sey45} 
   Seyfert, C. K., \& Nassau, J. J.  1945, \apj, 102, 377
\bibitem[Strom (1977)]{strom77} Strom, K. M. 1977,
  Standard Stars for IIDS Observations, Kitt Peak National Observatory
\bibitem[Tonry \& Davis (1979)]{ton79} 
   Tonry, J. T., \& Davis, M. 1979, \aj, 84, 1511
\bibitem[Valdes (1995)]{val95} 
   Valdes, F. 1995, Guide to the HYDRA Reduction Task DOHYDRA
\bibitem[van den Bergh \& Mackey (2004)]{vdB04} van den Bergh, S., 
  \& Mackey, A. D. 2004, MNRAS, 354, 713
\bibitem[Vete\v{s}nik (1962)]{vet62} 
   Vete\v{s}nik, M. 1962, Bull. Astr. Inst. Czech, 13, 180
\bibitem[Wirth, Smarr, \& Bruno (1985)]{wir85} 
   Wirth, A., Smarr, L. L., \& Bruno, T. L. 1985, \apj, 290, 140
\end{thebibliography}
\end{document}